\documentclass[a4paper,11pt]{article}
\usepackage{jheppub}

\usepackage[utf8]{inputenc}
\usepackage{amsmath}
\usepackage{graphicx}
\usepackage{multirow}

\newcommand{\orcid}[1]{\href{https://orcid.org/#1}{#1}}
\newcommand{\e}[1]{\times10^{#1}}

\newcommand{\Dmsqee}{\Delta m^2_{ee}}

\begin{document}
\preprint{CERN-TH-2023-015}
\title{\href{https://www.youtube.com/watch?v=KQetemT1sWc}{\textcolor{black}{Here Comes the Sun}}:\\Solar Parameters in Long-Baseline Accelerator Neutrino Oscillations}

\author[1]{Peter B.~Denton\note{\orcid{0000-0002-5209-872X}}}

\author[2,1]{and Julia Gehrlein\note{\orcid{0000-0002-1235-0505}}}
\affiliation[1]{High Energy Theory Group, Physics Department, Brookhaven National Laboratory, Upton, NY 11973, USA}
\affiliation[2]{Theoretical Physics Department, CERN,
             1 Esplanade des Particules, 1211 Geneva 23, Switzerland}

\emailAdd{pdenton@bnl.gov}
\emailAdd{julia.gehrlein@cern.ch}

\makeatletter
\hypersetup{colorlinks=true,allcolors=[rgb]{1,0.56,0},pdftitle=Here Comes the Sun: Solar Parameters in Long-Baseline Accelerator Neutrino Oscillations}
\makeatother

\abstract{
Long-baseline (LBL) accelerator neutrino oscillation experiments, such as NOvA and T2K in the current generation, and DUNE-LBL and HK-LBL in the coming years, will measure the remaining unknown oscillation parameters with excellent precision.
These analyses assume external input on the so-called ``solar parameters,'' $\theta_{12}$ and $\Delta m^2_{21}$, from solar experiments such as SNO, SK, and Borexino, as well as reactor experiments like KamLAND.
Here we investigate their role in long-baseline experiments.
We show that, without external input on $\Delta m^2_{21}$ and $\theta_{12}$, the sensitivity to detecting and quantifying CP violation is significantly, but not entirely, reduced.
Thus long-baseline accelerator experiments can actually determine $\Delta m^2_{21}$ and $\theta_{12}$, and thus all six oscillation parameters, without input from \emph{any} other oscillation experiment.
In particular, $\Delta m^2_{21}$ can be determined; thus DUNE-LBL and HK-LBL can measure both the solar and atmospheric mass splittings in their long-baseline analyses alone.
While their sensitivities are not competitive with existing constraints, they are very orthogonal probes of solar parameters and provide a key consistency check of a less probed sector of the three-flavor oscillation picture.
Furthermore, we also show that the true values of $\Delta m^2_{21}$ and $\theta_{12}$ play an important role in the sensitivity of other oscillation parameters such as the CP violating phase $\delta$.}

\date{\today}

\maketitle

\section{Long-baseline accelerator neutrino physics introduction}
Determining the six standard three-flavor oscillation parameters has been a top priority in the particle physics community since the discovery that they were physical in 1998 \cite{Super-Kamiokande:1998kpq}.
To date, remarkable progress has been made on several of the parameters.
In particular, $\theta_{13}$, $|\Delta m^2_{31}|$, $\theta_{12}$, and $\Delta m^2_{21}$ have all been determined to good precision.
The sign of $\Delta m^2_{31}$ is still to be determined, whether $\theta_{23}$ is in the upper octant, lower octant, or very close to maximal is an open question, and the complex CP violating (CPV) phase $\delta$ is largely unconstrained, see \cite{Denton:2022een} for a recent review.
The best experiments to probe these remaining unknowns are appearance experiments which are accomplished with long-baseline (LBL) accelerator\footnote{In this paper, LBL will refer only to accelerator experiments and not to long-baseline reactor experiments such as KamLAND or JUNO.} neutrinos in both neutrino mode and anti-neutrino mode.
While there are numerous partial degeneracies among these parameters, the ongoing experiments, currently running NOvA \cite{NOvA:2007rmc} and T2K \cite{T2K:2011qtm}, still have some discriminating capabilities among these parameters.
It is well established that the successors to these experiments, DUNE-LBL \cite{DUNE:2015lol} and HK-LBL \cite{Hyper-KamiokandeProto-:2015xww}, will have excellent precision to all three of the remaining unknowns with more than $5\sigma$ sensitivity to disfavor $\sin\delta=0$ for much of the parameter space.

Long-baseline oscillation analyses assume input from other experiments, however, in particular for the so-called ``solar parameters''\footnote{These parameters are referred to as solar parameters as they were first determined from solar neutrino data, but are now partially determined from solar data and partially from reactor neutrino data.
In the future the best constraints on these two fundamental parameters will be from reactor neutrino data.}: $\theta_{12}$ and $\Delta m^2_{21}$.
Many also include input on $\theta_{13}$ from medium baseline reactor experiments such as Daya Bay \cite{DayaBay:2018yms}, RENO \cite{RENO:2018dro}, and Double Chooz \cite{DoubleChooz:2019qbj}, although some long-baseline accelerator experiments will have comparable (within a factor of $\sim2$) sensitivity to this quantity \cite{DUNE:2020ypp}.

While it has been appreciated that a non-trivial three-flavor oscillation scenario is a necessary requirement for CP violation \cite{Cabibbo:1977nk,Jarlskog:1985ht,Krastev:1988yu}, a modern study on the impact of each of the other oscillation parameters on the final parameter, $\delta$, does not exist.
Specifically, it is important to understand the interplay of all of the oscillation parameters considering their now approximately known sizes  together with the fact that the final parameters will be measured in experiments experiencing the matter effect.

In this paper, we will show that input on the solar parameters $\Delta m^2_{21}$ and $\theta_{12}$ from other experiments is absolutely necessary to reach the physics goals of long-baseline accelerator experiments.
Then, we will investigate the sensitivity long-baseline accelerator experiments have to both their primary physics parameters, such as $\delta$, without input from $\Delta m^2_{21}$ and $\theta_{12}$, as well as the ability of long-baseline accelerator experiments to actually determine $\Delta m^2_{21}$ and $\theta_{12}$.
We will show that, without priors on $\Delta m^2_{21}$ and $\theta_{12}$ from solar and reactor experiments, the sensitivity to $\delta$ is significantly reduced due to some unusual oscillation scenarios where all the oscillation parameters take values very far from known values in an attempt to find agreement with the simulated data.
We will carefully investigate how this sensitivity depends on the precision on $\Delta m^2_{21}$ and $\theta_{12}$ (very little) and on the true value of $\Delta m^2_{21}$ and $\theta_{12}$ (modest dependence).
Then, since the sensitivity to determine $\delta$ does not go exactly to zero, this means that DUNE-LBL and HK-LBL will have some sensitivity to measure $\Delta m^2_{21}$ and $\theta_{12}$ in their long-baseline channels; we will determine the statistical level at which the long-baseline accelerator experiments can actually determine $\Delta m^2_{21}$ and $\theta_{12}$.

Other studies exist exploring the impact of $\Delta m^2_{21}$ and $\theta_{12}$ in experiments not traditionally designed to measure these parameters.
For example, \cite{Seo:2018rrb,Hernandez-Cabezudo:2019qko} investigated the ability to probe these parameters at Daya Bay and RENO where the $\Delta m^2_{21}$ oscillations have only just started to develop.
They found that these experiments can constraint $|\Delta m^2_{21}|\lesssim20\e{-5}$ eV$^2$.
In addition, \cite{Capozzi:2020cxm,Forero:2021lax} found that the true values of $\Delta m^2_{21}$ and $\theta_{12}$ within their current uncertainties have a potentially sizable impact on JUNO's sensitivity to the sign of $\Delta m^2_{31}$.

Throughout the paper we will show results for characteristic experiments or parameters that best highlights the physics.
Other combinations of the results are shown in the appendix \ref{sec:add_fig} for completeness.
We begin the manuscript by providing an analytical understanding of the impact of $\Delta m^2_{21}$ and $\theta_{12}$ on the measurement of CPV in sec.~\ref{sec:solar overview} followed by a description of our numerical analysis in sec.~\ref{sec:analysis}. We present our results in sec.~\ref{sec:results} and then discuss them and conclude in sec. \ref{sec:discussion}. In appendix \ref{sec:other_para} we demonstrate the 
impact of $\Delta m^2_{21}$ and $\theta_{12}$ on the determination of $\theta_{23},~\theta_{13}$, and $\Delta m_{31}^2$.

\section{The role of the solar parameters in the CPV measurement}
\label{sec:solar overview}
All three mixing angles need to be non-zero to allow for CPV in the neutrino sector \cite{Cabibbo:1977nk,Jarlskog:1985ht,Krastev:1988yu}. Furthermore, their values together with the value of $\delta$ dictate the size of CPV in the lepton sector, measured via the Jarlskog invariant $J=s_{12}c_{12}s_{13}c_{13}^2s_{23}c_{23}\sin\delta$ \cite{Jarlskog:1985ht} where we use the common notation $s_{ij}\equiv\sin\theta_{ij}$ and $c_{ij}\equiv\cos\theta_{ij}$. Hence a measurement of all oscillation parameters is required to quantify leptonic CPV. While $\theta_{13}$ is already well measured and the interplay between a more precise measurement of $\theta_{23}$ and $\delta$ at future LBL experiments has been studied before \cite{Coloma:2012wq,Minakata:2013eoa,Agarwalla:2013ju,Coloma:2014kca,Ghosh:2015ena,Agarwalla:2022xdo}, the role of the solar parameters, $\Delta m^2_{21}$ and $\theta_{12}$, in the leptonic CPV measurement has not been analysed in detail.
In the following we will 
therefore conduct a detailed study of the role of $\Delta m^2_{21}$ and $\theta_{12}$ at LBL accelerator experiments.

The two solar parameters, $\theta_{12}$ and $\Delta m^2_{21}$, have been determined in solar experiments such as SNO \cite{SNO:2002tuh}, SK \cite{Super-Kamiokande:2016yck}, Borexino \cite{Borexino:2017rsf,BOREXINO:2018ohr}, Homestake \cite{Cleveland:1998nv}, GALLEX \cite{Kaether:2010ag}, and SAGE \cite{SAGE:2009eeu}.
The values of those parameters have been confirmed in the long-baseline reactor experiment KamLAND \cite{KamLAND:2013rgu}.
In particular, a combined fit of solar data provides a good measurement of $\theta_{12}$  and KamLAND's reactor measurement of $\theta_{12}$ is only a bit less constraining.
The constraint on the frequency, $\Delta m^2_{21}$, is dominated by KamLAND with some additional information from solar data, albeit at significantly lower precision.
A small tension briefly existed between solar and reactor determinations of $\Delta m^2_{21}$ at the $\sim2\sigma$ level \cite{Maltoni:2015kca,Liao:2017awz}, although this seems to have evaporated with new solar data and analyses from SK \cite{yusuke_koshio_2022}.

Nevertheless, an ambiguity exists in the definition of the $\Delta m^2_{21}$ and $\theta_{12}$, and really in the definition of all three mass states.
Multiple viable definitions exist, see e.g.~\cite{Denton:2020exu,Denton:2021vtf}.
One possible definition is $m_1<m_2<m_3$, although until the atmospheric mass ordering is known, this leads to rather complicated conditional expressions for many oscillation experiments.
Another possible definition is
\begin{equation}
|U_{e1}|>|U_{e2}|>|U_{e3}|\,,
\label{eq:mass state defn}
\end{equation}
which is this definition that we choose to use in the following.
We use this definition since we know the magnitude of all three elements of the electron neutrino row quite well from medium- and long-baseline reactor neutrino experiments, as well as solar experiments.
This definition means that $\theta_{12}<45^\circ$, $\sin\theta_{12}>\tan\theta_{13}$, while $\Delta m^2_{21}$ can be positive or negative.
We note that the definition in eq.~\ref{eq:mass state defn} differs from another definition that is sometimes used which is: $m_1<m_2$, $|U_{e3}|<|U_{e1}|$, and $|U_{e3}|<|U_{e2}|$ which means that $\Delta m^2_{21}>0$ and $\tan\theta_{13}<\min(\sin\theta_{12},\cos\theta_{12})$.
Thus the practical difference between these two definitions is that the fact that the $^8$B solar neutrino disappearance probability is $P_{ee}^{^8B}\sim\frac13$ tells us that $\Delta m^2_{21}>0$ in our definition, while in the other definition it tells us that $\theta_{12}<45^\circ$.

Solar parameters have some partial degeneracies with the CP phase as well as some other parameters.
For example, in vacuum near the first oscillation maximum when $\Delta m^2_{32}L/4E\simeq\pi/2$ the CP difference is
\begin{align}
P(\nu_\mu\to\nu_e)-P(\bar\nu_\mu\to\bar\nu_e)&=-16J\sin\left(\frac{\Delta m^2_{31}L}{4E}\right)\sin\left(\frac{\Delta m^2_{32}L}{4E}\right)\sin\left(\frac{\Delta m^2_{21}L}{4E}\right)\,,\label{eq:cpasym long}\\
&\approx -8\pi J\frac{\Delta m^2_{21}}{\Delta m^2_{32}}\,,
\label{eq:cpasym short}
\end{align}
where $J$ is the Jarlskog invariant.
Thus without knowledge of $\Delta m^2_{21}$ or $\theta_{12}$ there is a degeneracy between $\sin\delta$ and the solar parameters, up to the limit from unitarity $|J|\le\frac1{6\sqrt3}\approx0.096$.
We note, however, that in vacuum there is no asymmetry if CP is conserved and it is impossible to ``dial up'' $\Delta m^2_{21}$ and $\sin2\theta_{12}$ enough to get something that looks like CP violation.
Equations \ref{eq:cpasym long}-\ref{eq:cpasym short} also highlight the important role a non-zero value of $\Delta m^2_{21}$ plays in vacuum oscillations.
That is, all three mass states must be different in order to have CP violation.
This can be seen in other ways as well in that if $m_1=m_2$ then the mixing angle $\theta_{12}$ is no longer physical which also removes the possibility to detect CP violation.

For DUNE-LBL and HK-LBL, however, the matter effect plays a role in oscillations.
Among other things, this leads to an \emph{apparent} CP violating effect
\cite{Kuo:1987km,Krastev:1988yu,Toshev:1989vz,Tanimoto:1996ky,Arafune:1996bt,Arafune:1997hd,Bilenky:1997dd,Minakata:1998bf,Koike:1999tb,Barenboim:1999mp,Parke:2000hu} with the same $(L/E)^3$ dependence in eq.~\ref{eq:cpasym long}.
The probability does not, however, depend on $\delta$ if $\Delta m^2_{21}\to0$, as we outline here.
The matter equivalent version of $\Delta m^2_{21}$, denoted with a hat as $\Delta\widehat{m^2}_{21}$, is always non-zero even when $\Delta m^2_{21}=0$ and is well approximated \cite{Denton:2016wmg,Denton:2018hal} (see also \cite{DeRujula:2000ap}) as
\begin{equation}
\lim_{\Delta m^2_{21}\to0}\Delta\widehat{m^2}_{21}\approx a\cos^2\hat\theta_{13}+\Dmsqee\sin^2(\hat\theta_{13}-\theta_{13})\,,
\end{equation}
where $a=2\sqrt2G_FN_eE$ is the contribution from the matter effect, $N_e$ is the electron density, $E$ is the neutrino energy, $\Dmsqee=\cos^2\theta_{12}\Delta m^2_{31}+\sin^2\theta_{12}\Delta m^2_{32}$ \cite{Nunokawa:2005nx,Parke:2016joa}, and
\begin{equation}
\cos2\hat\theta_{13}\approx\frac{\cos2\theta_{13}-a/\Dmsqee}{(\cos2\theta_{13}-a/\Dmsqee)^2+\sin^22\theta_{13}}\,.
\end{equation}
Therefore it appears as though this will nonetheless lead to apparent CPV that still depends on $\delta$ and has the same $(L/E)^3$ dependence as in eq.~\ref{eq:cpasym long}.
However, we must account for the behavior of the Jarlskog coefficient in matter.
From \cite{Denton:2019yiw} we have that
\begin{equation}
\hat J\approx\frac J{\sqrt{(\cos2\theta_{12}-c_{13}^2a/\Delta m^2_{21})^2+\sin^22\theta_{12}}\sqrt{(\cos2\theta_{13}-a/\Delta m^2_{ee})^2+\sin^22\theta_{13}}}\,,
\end{equation}
where the corrections to this approximation are proportional to $\Delta m^2_{21}$, thus it becomes exact at $\Delta m^2_{21}\to0$.
Thus $\hat J\to0$ as $\Delta m^2_{21}\to0$ and therefore the triple sine term is zero in this limit in matter as well.
In fact, due to the Naumov-Harrison-Scott identity \cite{Naumov:1991ju,Harrison:1999df},
\begin{equation}
\hat J\Delta\widehat{m^2_{32}}\Delta\widehat{m^2_{31}}\Delta\widehat{m^2_{21}}=J\Delta m^2_{32}\Delta m^2_{31}\Delta m^2_{21}\,,
\end{equation}
and the fact that none of the $\Delta\widehat{m^2_{ij}}\to0$ as $\Delta m^2_{ij}\to0$, $\hat J\to0$ if any of the three $\Delta m^2_{ij}\to0$.
That is, even though the effective mass squared splitting is always non-zero in matter, there is no impact due to real CPV in neutrino oscillations if $\Delta m^2_{21}=0$.

In addition to the role the matter effect plays, the simple story shown in eq.~\ref{eq:cpasym short} is further complicated by several additional effects.
First, via the presence of a near detector and a careful understanding of the flux and cross sections, neutrino oscillation experiments measure each appearance channel, $P(\nu_\mu\to\nu_e)$ and $P(\bar\nu_\mu\to\bar\nu_e)$, independently.
Second, DUNE-LBL -- and to a lesser extent HK-LBL -- measure the spectrum around the oscillation maximum in the appearance channels.
This provides key shape information.
Third, it is conceivable that DUNE-LBL and/or HK-LBL might be able to gain some information about the second oscillation maximum which would provide additional important information about CP violation.
Finally, the matter effect \cite{Wolfenstein:1977ue} which is quite important for DUNE-LBL and NOvA, plays a key role as discussed above.
In particular it eases the measurement of the atmospheric matter effect which reduces a key degeneracy for DUNE-LBL and, to a lesser extent, HK-LBL.

\begin{figure}
\centering
\includegraphics[width=0.8\textwidth]{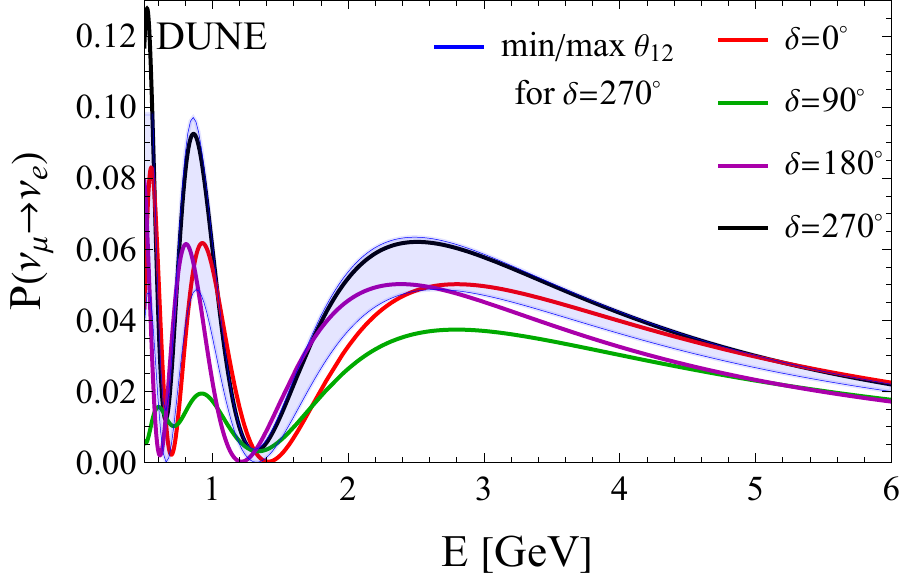}
\includegraphics[width=0.8\textwidth]{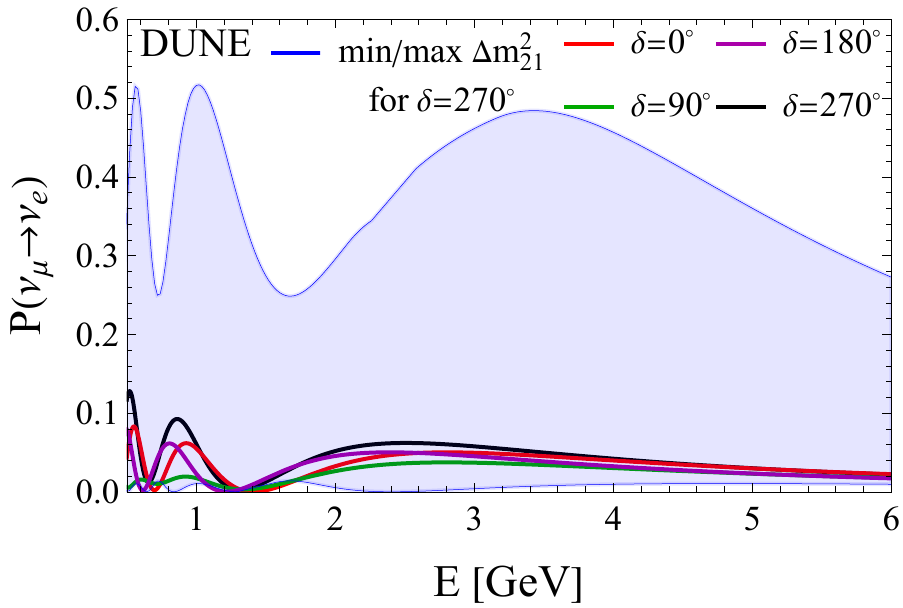}
\caption{The oscillation probability at DUNE-LBL for $\delta=0,~90^\circ,~180^\circ,~270^\circ$ and fixing the oscillation parameters to the benchmark scenario defined in tab.~\ref{tab:solar bf}  and in the text as a function of energy are shown as colored lines. The blue regions show the extreme values of the   probability assuming $\delta=-90^\circ$  varying either $\theta_{12}$ between 0 and $45^\circ$ (top) or $\Delta m_{21}^2$ between $-\Delta m_{31}^2$ and $\Delta m_{31}^2$ (bottom). If the colored lines are contained in the blue regions $\delta=-90^\circ$ cannot be easily distinguished from other values of $\delta$ without a prior on the solar parameter.}
\label{fig:osciprob}
\end{figure}

In fig.~\ref{fig:osciprob}
we illustrate the impact of $\Delta m^2_{21}$ and $\theta_{12}$ on the appearance probability at DUNE-LBL to set the stage for the rest of the paper. We use a matter density of $\rho=3~\text{g/cc}$, a baseline of 1300 km, and as benchmark the oscillation parameters defined in the next section and in tab.~\ref{tab:solar bf} and $\delta=-90^\circ$. We choose $\Delta m^2_{21}$ and $\theta_{12}$ which extremize the probability as a function of the energy to arrive at a possible range  of probabilities. To demonstrate the effects of each of $\Delta m^2_{21}$ and $\theta_{12}$, we vary only one of them and  allow $\theta_{12}\in [0,~45^\circ]$ and $\Delta m_{21}^2\in[-\Delta m_{31}^2,\Delta m_{31}^2]$ to ensure that the two mass splittings remain different. This envelop can be compared to the probabilities with $\delta=0,~90^\circ,~180^\circ,~270^\circ$ and fixing $\Delta m^2_{21}$ and $\theta_{12}$ to the SK+SNO+KamLAND best fit. If these probabilities are enclosed in the envelop of probabilities with extreme values of $\Delta m^2_{21}$ and $\theta_{12}$, $\delta=-90^\circ$ may not be easily distinguishable from other values of $\delta$ without the addition  of solar priors.
We find as  extreme values for $\theta_{12}$ at the peak of the DUNE neutrino flux  at $E=3$ GeV $\theta_{12}^{min}\approx 2.7^\circ$, $\theta_{12}^{max}\approx 44^\circ$, and for $\Delta m_{21}^2$ at $E=3$ GeV $\Delta m_{21}^{2,min}\approx -5.6\times 10^{-4}~\text{eV}^2,~\Delta m_{21}^{2,min}\approx 2.3\times 10^{-3}~\text{eV}^2$.
From the upper plot of fig.~\ref{fig:osciprob} we see that even with extreme values of $\theta_{12}$ the changes of the oscillation amplitude are not dramatic and some probability curves with fixed $\Delta m^2_{21}$ and $\theta_{12}$ lay outside of the envelop. On the other hand, from the lower plot of fig.~\ref{fig:osciprob}, we see that the effects of changes in $\Delta m_{21}^2$ are more pronounced and all probabilities with fixed $\Delta m^2_{21}$ and $\theta_{12}$ are contained in the envelop.
Note, however, that fig.~\ref{fig:osciprob} does not provide any shape information about the behavior at the first oscillation maximum as the parameters are varied, as well as any potential impact at the second oscillation maximum.
We therefore conclude that it seems likely that priors on both $\Delta m^2_{21}$ and $\theta_{12}$ are important to obtain sensitivity to $\delta$ and to achieve precision on $\delta$.

\section{Analysis details}
\label{sec:analysis}
To estimate the sensitivities at various LBL experiments we use the GLoBES software package \cite{Huber:2004ka}.
We use the publicly available experimental files for NOvA \cite{NOvA:2004blv,Yang_2004}, T2K \cite{Huber:2002mx,T2K:2001wmr,Ishitsuka:2005qi}, DUNE \cite{DUNE:2021cuw}, and HK-LBL \cite{Huber:2002mx,T2K:2001wmr,Ishitsuka:2005qi} and modify them to ensure agreement to the most recent quoted sensitivities from the experiments for a given set of assumptions about the oscillation parameters \cite{t2hkichep,novawin,t2ktaup}.
For each of the four long-baseline accelerator experiments, we consider both neutrino mode and anti-neutrino mode as well as both disappearance ($P(\nu_\mu\to\nu_\mu)$) and appearance ($P(\nu_\mu\to\nu_e)$) modes.
We do not include $\nu_\tau$ appearance mode which may be relevant for DUNE-LBL \cite{DeGouvea:2019kea,Machado:2020yxl,Kosc:2021huh}, see also \cite{Abraham:2022jse}.
The experimental details of all four long-baseline accelerator experiments are summarized in table \ref{tab:exps} and are set to match the latest experimental sensitivity curves for the assumed oscillation parameters. 
While these details may change as the upcoming experiments evolve, we have checked that they do capture the relevant features and that changes in the exposure do not significantly modify the results.

\begin{table}
\centering
\caption{A summary of the relevant experimental details assumed for each experiment where POT is the total accumulated protons on target and $\nu$:$\bar\nu$  is the ratio  of neutrino to anti-neutrino mode.}
\begin{tabular}{c|c|c|c|c}
Experiment & Technology & Fiducial Volume & Total POT ($\nu$+$\bar\nu)$ & $\nu$:$\bar\nu$ \\\hline\hline
NOvA & Scintillator &25 kT& $7.2\times 10^{21}$&1:1 \\
T2K & Water Cherenkov &22.5 kT&$10\times 10^{21}$& 1:1\\\hline
DUNE-LBL & LArTPC &40 kT&$14\times 10^{21}$&1:1 \\
HK-LBL & Water Cherenkov &190 kT&$27\times 10^{21}$&1:3
\end{tabular}
\label{tab:exps}
\end{table}

\begin{figure}
\centering
\includegraphics[width=0.8\textwidth]{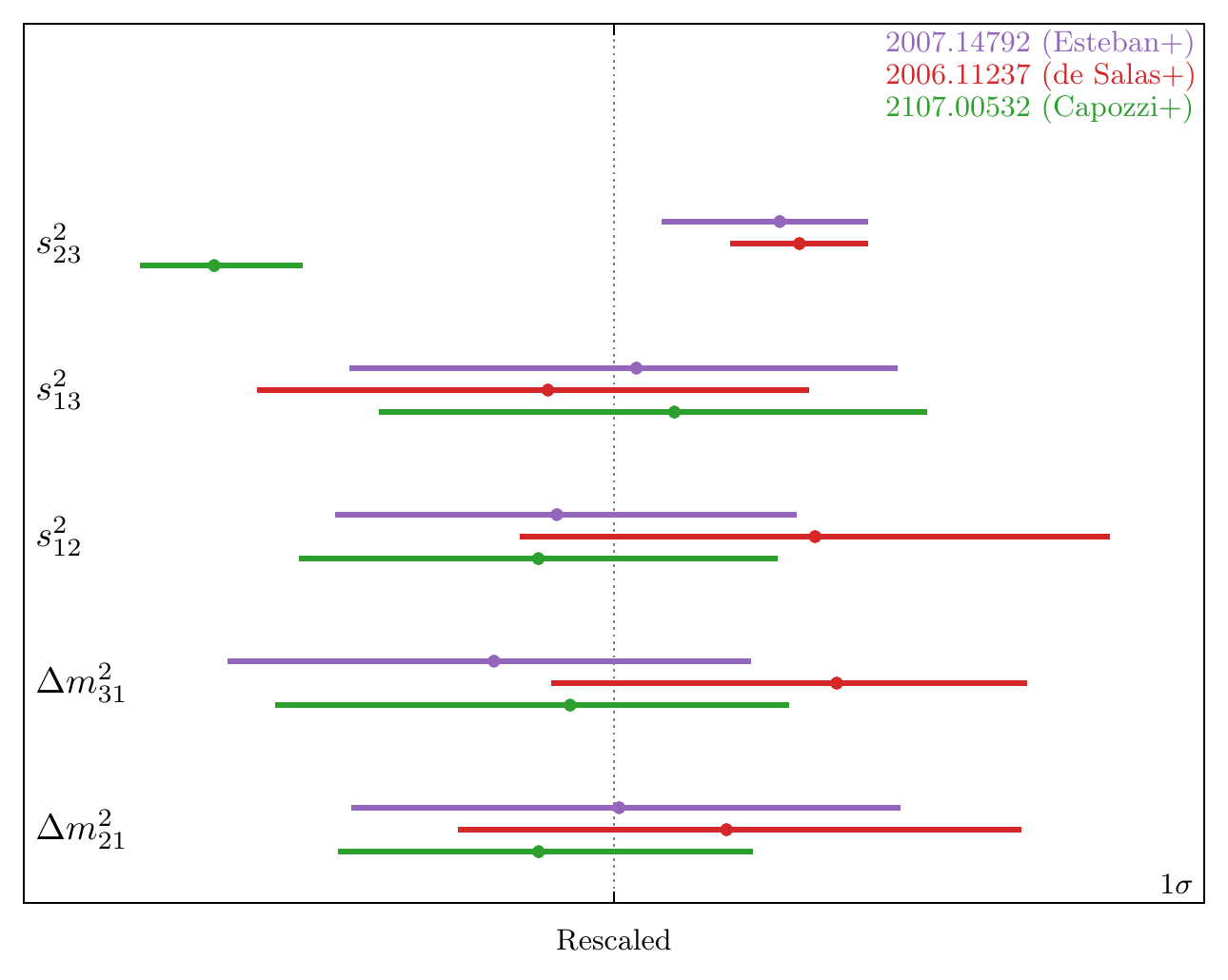}
\caption{The comparison on the preferred values of the five oscillation parameters (not including $\delta$) from the three primary global fits \cite{deSalas:2020pgw,Esteban:2020cvm,Capozzi:2021fjo}.
The quoted $1\sigma$ uncertainties are shown and the normal ordering is assumed.
The vertical dashed line is at the weighted average and the spread is rescaled to fit on the same scale; $s_{23}^2$ is rescaled $2.5\times$ as much as the other parameters.}
\label{fig:gf compare}
\end{figure}

To study the sensitivity to the oscillation parameters we make various assumptions on the priors of $\Delta m^2_{21}$ and $\theta_{12}$.
The case of no priors  provides a testament to what an experiment can do entirely on their own.
Our current knowledge of $\Delta m^2_{21}$ and $\theta_{12}$ comes from solar data, KamLAND, and a combined analysis of both; the last of these is the closest approximation to the fiducial analyses that most experiments run. This is also the benchmark scenario we will use in the following, unless otherwise stated.
In the future our knowledge of $\Delta m^2_{21}$ and $\theta_{12}$ will increase with information from solar neutrinos at HK-LBL or DUNE-LBL and information from reactor neutrinos at JUNO\footnote{JUNO will also have sensitivity to $\Delta m^2_{21}$ and $\theta_{12}$ via solar neutrinos \cite{JUNO:2022jkf}, but will not be competitive with DUNE-solar.}.
An overview of our current knowledge on $\Delta m^2_{21}$ and $\theta_{12}$ can be found in tabs.~\ref{tab:solar bf}, \ref{tab:priors} which also includes the global fit results from
\cite{deSalas:2020pgw,Esteban:2020cvm,Capozzi:2021fjo}. We see that the best fit values of $\Delta m^2_{21}$ and $\theta_{12}$ vary among the different determinations by a $\sim1\sigma$ spread among the global fits. This is illustrated in fig.~\ref{fig:gf compare} where we also include global fit results on the remaining oscillation  parameters (not including $\delta$).
This figure shows that, in fact, there is a $\sim1\sigma$ difference among the global fits for many of the parameters including $\Delta m^2_{21}$ and $\theta_{12}$.

\begin{table}
\centering
\caption{The current best fit values for $\Delta m_{21}^2$ and $\theta_{12}$ including different data sets.
Unless otherwise specified, the bolded values are the default values taken.}
\begin{tabular}{c|c|c|c}
Data & $\Delta m^2_{21}$ [$10^{-5}$ eV$^2$] & $\sin^2\theta_{12}$ & Ref.\\\hline
SK+SNO & 6.10 & 0.305 & \cite{yusuke_koshio_2022}\\
KamLAND & $\pm7.54$ & 0.316 & \cite{KamLAND:2013rgu}\\
\textbf{SK+SNO+KamLAND} & \textbf{7.49} & \textbf{0.305} & \cite{yusuke_koshio_2022}\\\hline
\multirow{3}{*}{Global fit} & 7.42 & 0.304 & \cite{Esteban:2020cvm}\\
 & 7.5 & 0.318 & \cite{deSalas:2020pgw}\\
 & 7.36 & 0.303 & \cite{Capozzi:2021fjo}
\end{tabular}
\label{tab:solar bf}
\end{table}

\begin{table}
\centering
\caption{The precision on $\Delta m^2_{21}$ and $\theta_{12}$ used in different cases and the associated reference for the precise input used.
Unless otherwise specified, the bolded values are the default values taken.
HK will also measure solar neutrinos \cite{Hyper-KamiokandeProto-:2015xww} but with a precision comparable to the total current solar data \cite{Martinez-Mirave:2021cvh} of 14\% and 5.6\% for $\Delta m^2_{21}$ and $\sin^2\theta_{12}$ respectively.}
\begin{tabular}{c|l|c|c|c}
 \multicolumn{1}{c}{} & \multicolumn{1}{c}{} & \multicolumn{2}{c}{$\delta x/x$}\\
Generation & Data & $\Delta m^2_{21}$ & $\sin^2\theta_{12}$ & Ref.\\\hline
\multirow{6}{*}{Current} & SK+SNO & 15\% & 4.6\% &\cite{yusuke_koshio_2022}\\
 & KamLAND & 2.5\% & 9.5\%&\cite{KamLAND:2013rgu}\\
 & \textbf{SK+SNO+KamLAND} & \textbf{2.4\%} & \textbf{4.3\%} & \cite{yusuke_koshio_2022}\\\cline{2-5}
 & \multirow{3}{*}{Global fit} & 2.8\% & 4.3\% & \cite{Esteban:2020cvm}\\
 & & 2.9\% & 5.0\% & \cite{deSalas:2020pgw}\\
 & & 2.2\% & 4.3\% & \cite{Capozzi:2021fjo}\\\hline
\multirow{2}{*}{Future} & DUNE-solar & 5.9\% & 3.0\%&\cite{Capozzi:2018dat}\\
 & JUNO & 0.3\% & 0.5\% &\cite{JUNO:2022mxj}
\end{tabular}
\label{tab:priors}
\end{table}

In our study, for the remaining parameters, when we use priors on them, we assume 
\begin{align}
\sin^2 2 \theta_{13}&= 0.0853~ (\pm 2.8\%) ~\text{from  \cite{kam_biu_luk_2022_6683712}},\nonumber\\
\Delta m_{32}^2&=2.454 \times 10^{-3}~\text{eV}^2 ~(\pm 2.3\%)~ \text{from  \cite{kam_biu_luk_2022_6683712}}~,\nonumber\\
\sin^2 \theta_{23}&= 0.57~(\pm 7.0\%) ~\text{from \cite{NOvA:2021nfi}}
\end{align}
where the first two parameters were determined from the most recent Daya Bay results \cite{kam_biu_luk_2022_6683712,DayaBay:2022orm}\footnote{The latest Daya Bay results \cite{DayaBay:2022orm} are $<1\sigma$ different from the numbers mentioned here.} and we use the current results from NOvA on $\sin^2\theta_{23}$ \cite{NOvA:2021nfi}.
The choice of priors on $\theta_{13}$, $\Delta m^2_{32}$, and $\sin^2\theta_{23}$ will not strongly affect our results as the future LBL experiments are able to determine these parameters on their own with good precision as indirectly shown in fig.~\ref{fig:mcdonalds subtract} below.

Finally, we assume the true mass ordering to be normal but we  test both orderings in our analysis, i.e.~we do not fix the mass ordering in our analysis.

\section{Results}
\label{sec:results}
In this section we present various numerical results to support our claims on the importance of solar neutrino parameters in long-baseline experiments.
We present the results for DUNE-LBL while similar conclusions can be reached for HK-LBL as well.
We also calculate the same results for NOvA and T2K, however their sensitivities are usually much less competitive such that we omitted them from plots.
Additional results are shown in appendix \ref{sec:add_fig}.

\subsection{Sensitivity to the complex phase}
\begin{figure}
\centering
\includegraphics[width=0.9\textwidth]{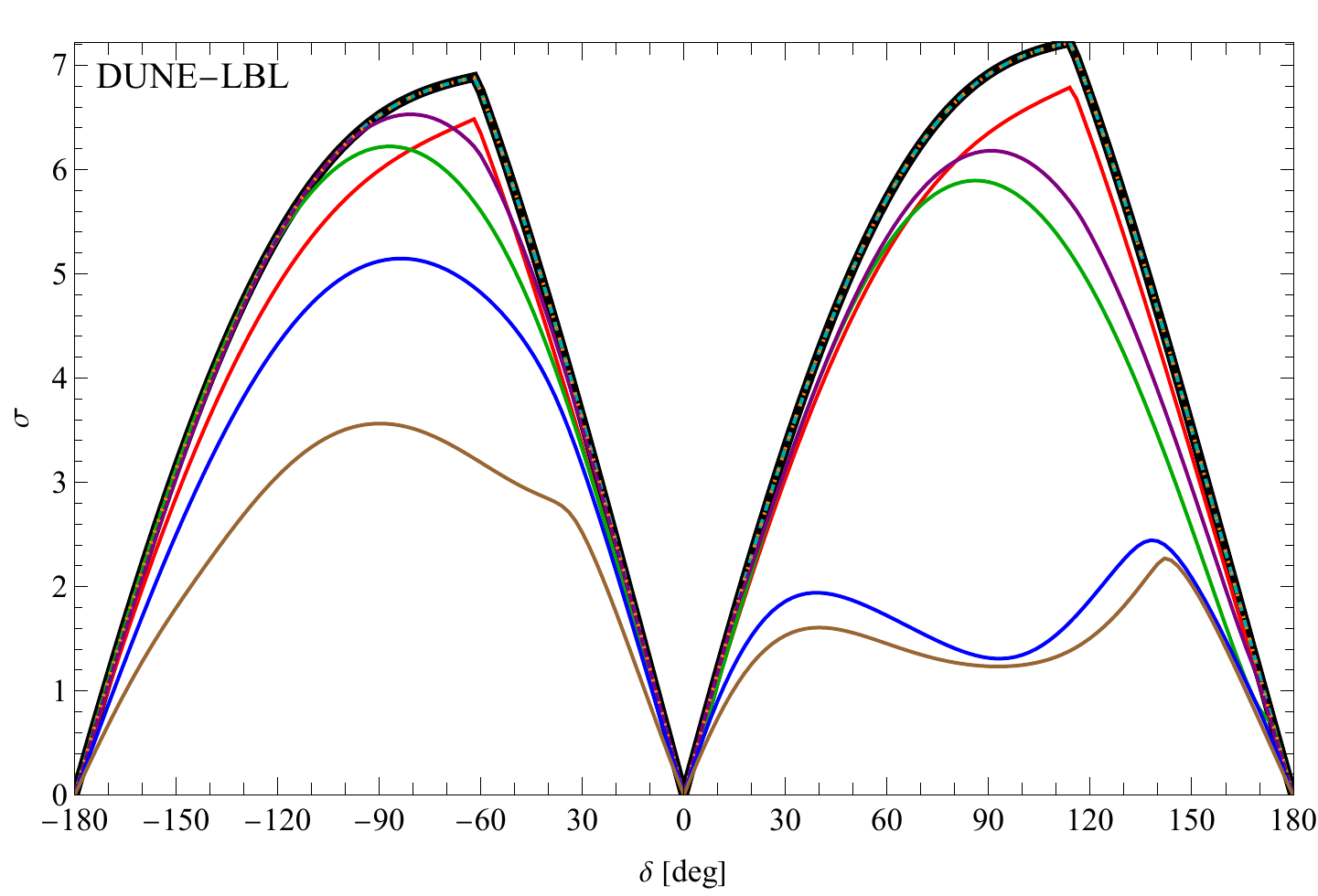}
\includegraphics[width=0.9\textwidth]{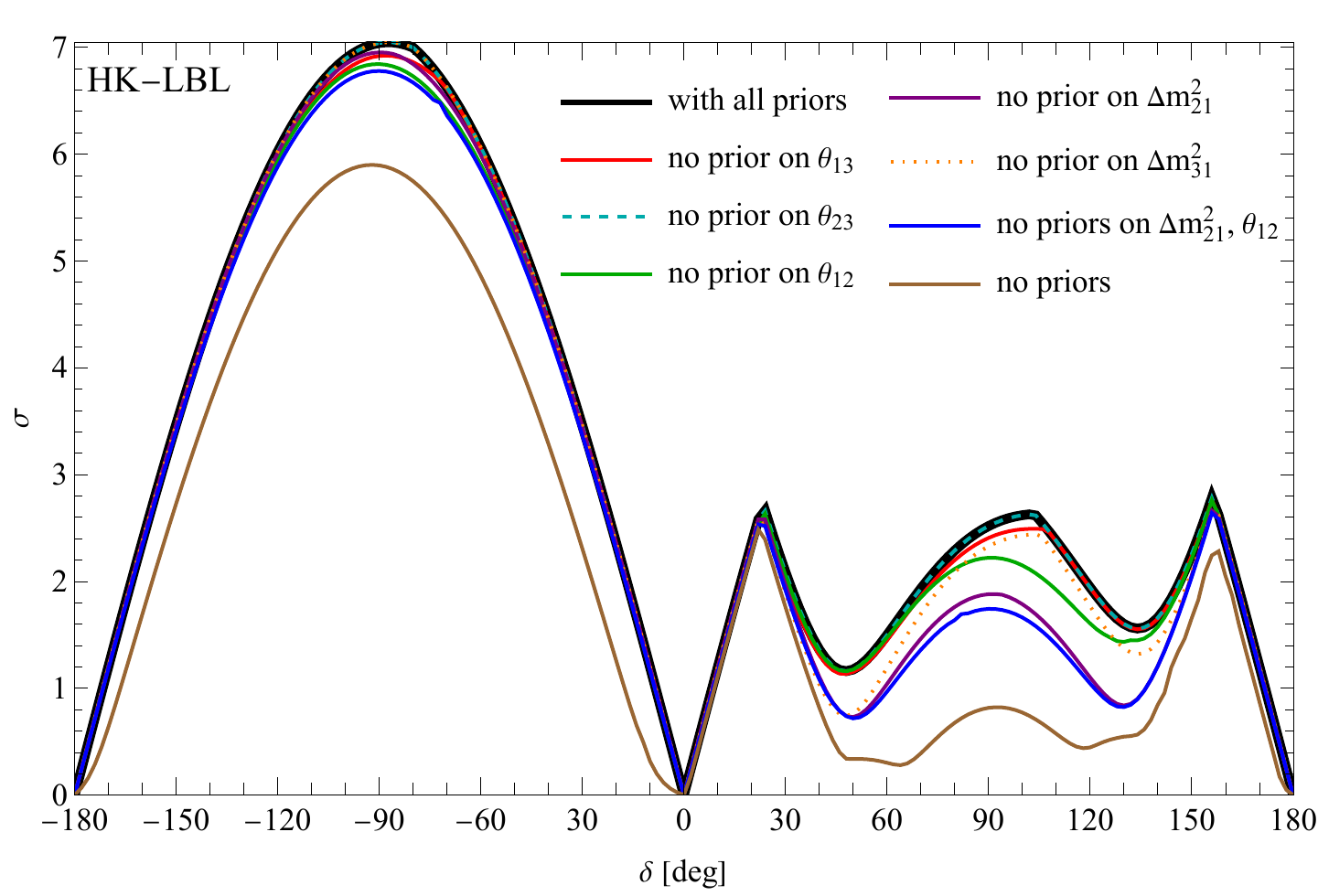}
\caption{The sensitivity of DUNE-LBL (top) and HK-LBL (bottom) to disfavor $\sin\delta=0$ in the NO as a function of the true value of $\delta$ using the benchmark scenario defined in tab.~\ref{tab:priors} and in the text.
In black is the sensitivity with a prior on all five constrained oscillation parameters while the other colors are the sensitivity with priors on all but one of the parameters.
For DUNE-LBL, the curves without a prior on $\theta_{23}$ or $\Delta m_{31}^2$ coincide with the curve assuming priors on all oscillation parameters due to the excellent sensitivity of DUNE-LBL to these parameters; the same is true for HK-LBL, but only for the $\theta_{23}$ curve.
The mass ordering is free.}
\label{fig:mcdonalds subtract}
\end{figure}

We first investigate the sensitivity to disfavor $\sin\delta=0$ for DUNE-LBL and HK-LBL with priors on all five oscillation parameters, $\theta_{12}$, $\theta_{13}$, $\theta_{23}$, $\Delta m^2_{21}$, and $\Delta m^2_{31}$, as described above.
The sensitivity, shown in black in both panels of fig.~\ref{fig:mcdonalds subtract} is in excellent agreement with DUNE-LBL's quoted sensitivity \cite{DUNE:2021cuw} and HK-LBL's sensitivity from \cite{t2hkichep}.
We then remove each of the five priors, one at a time to see which, if any, affects the sensitivity.
We see that removing the priors on $\theta_{23}$ and $\Delta m^2_{31}$ have little effect as expected since DUNE-LBL can provide an excellent measurement of these parameters without further input.
Removing the prior on $\theta_{13}$ has a small effect on the sensitivity to $\delta$.
Removing the priors on one of $\Delta m^2_{21}$ or $\theta_{12}$ is comparable in effect to removing the prior on $\theta_{13}$, while removing \emph{both} priors on $\Delta m^2_{21}$ and $\theta_{12}$ dramatically reduces the sensitivity to $\delta$, in particular for $\delta\in[0,180^\circ]$ where the sensitivity is at the $\sim2\sigma$ level at best; for $\delta\in[-180^\circ,0]$ the sensitivity is only at $5\sigma$ for $\delta$ very close to $-90^\circ$.
Finally, with no priors from other experiments, the sensitivity to $\delta$ is at best $\sim3.5\sigma$ at $\delta\simeq-90^\circ$.
This dramatic reduction in sensitivity comes at fairly unusual oscillation parameters, known to be dramatically inconsistent with other oscillation measurements, most notably $\Delta m^2_{21}$ up to $\sim60\e{-5}$ eV$^2$ and $\theta_{12}$ taking any value from 0 to $45^\circ$.

Similar conclusions about the importance of $\Delta m^2_{21}$ and $\theta_{12}$ can be reached for HK-LBL as well.
Unlike DUNE-LBL, HK-LBL cannot determine the mass ordering with high sensitivity, therefore the sensitivity to CPV in the range $\delta \in[0,180^\circ]$ is below $3\sigma$ even when including priors on all oscillation parameters. However also for HK-LBL a drop in the sensitivity arises from removing priors on both $\Delta m^2_{21}$ and $\theta_{12}$, in particular for $\delta\approx 45^\circ,~135^\circ$ where the sensitivity falls below $1\sigma$.
If we fix the mass ordering, then the reduction in sensitivity to $\delta$ without the inclusion of solar priors is less drastic but still appreciable as shown in fig.~\ref{fig:hk_seni}.

The current generation of LBL experiments, NOvA and T2K, have a CPV sensitivity of $\lesssim 2\sigma$ even when including all priors, see fig.~\ref{fig:hk_seni}.
Nevertheless, also for them we find a reduced CPV sensitivity in the absence of solar priors.
Similar to HK-LBL, T2K's CPV sensitivity for $\delta\in [0,180^\circ]$ is improved when fixing the mass ordering.

\begin{figure}
\centering
\includegraphics[width=0.9\textwidth]{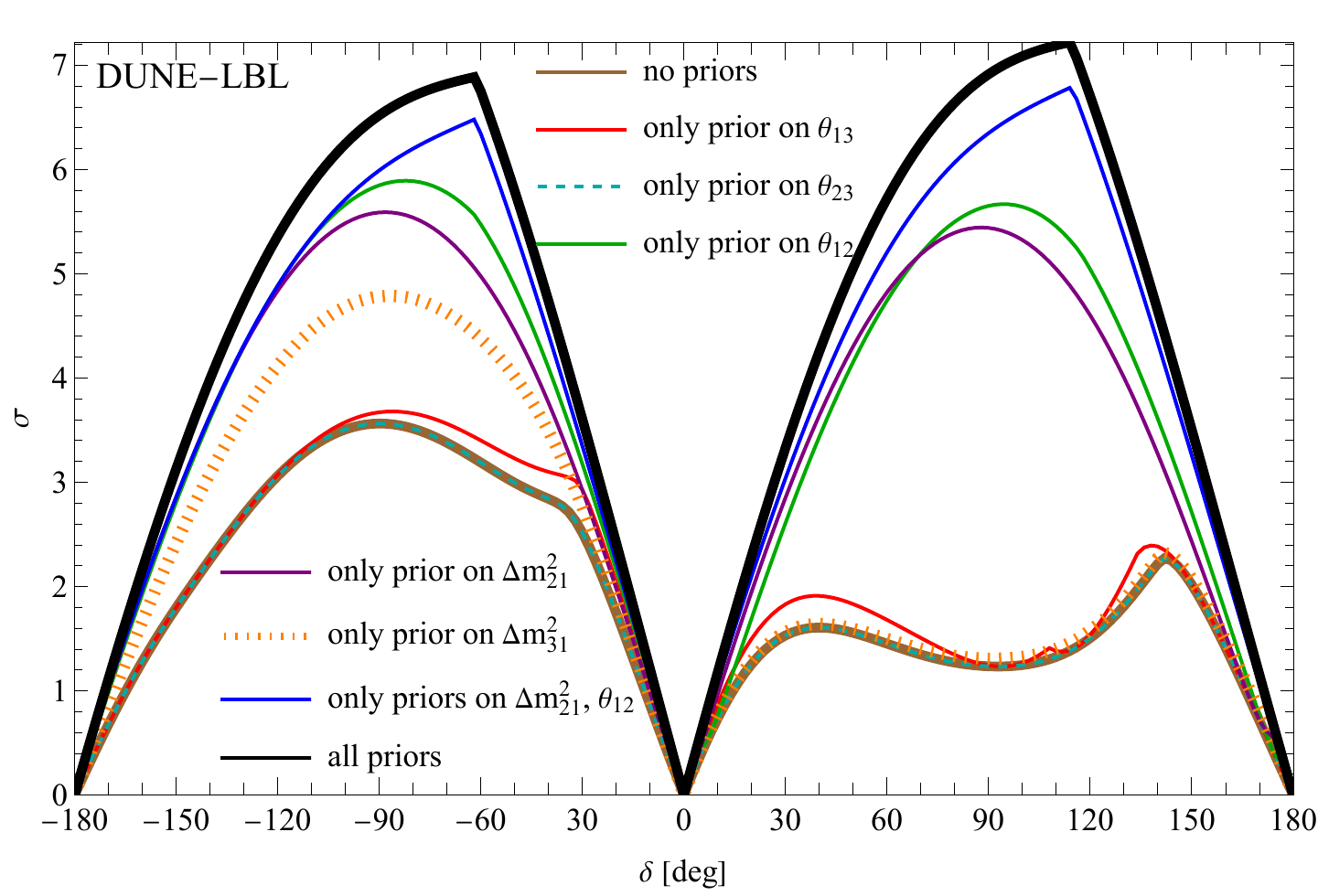}
\caption{The same as fig.~\ref{fig:mcdonalds subtract} at DUNE-LBL but now starting with no prior in brown and the other curves are with one prior on one parameter. The curve with only a prior on $\theta_{23}$ lays on top of the curve without priors.}
\label{fig:mcdonalds add}
\end{figure}

Next we show in fig.~\ref{fig:mcdonalds add} again the sensitivity at DUNE-LBL without priors.
We then include priors one at a time.
We see, consistent with the above text, the impact of priors on $\theta_{23}$ and $\Delta m^2_{31}$ are negligible and that including the prior on $\theta_{13}$ provides only a marginal improvement.
Including either of the solar priors significantly enhances the sensitivity to $>5\sigma$ for some values near $\delta\simeq\pm90^\circ$ and including both solar priors increases the sensitivity even more.

\begin{figure}
\centering
\includegraphics[width=0.9\textwidth]{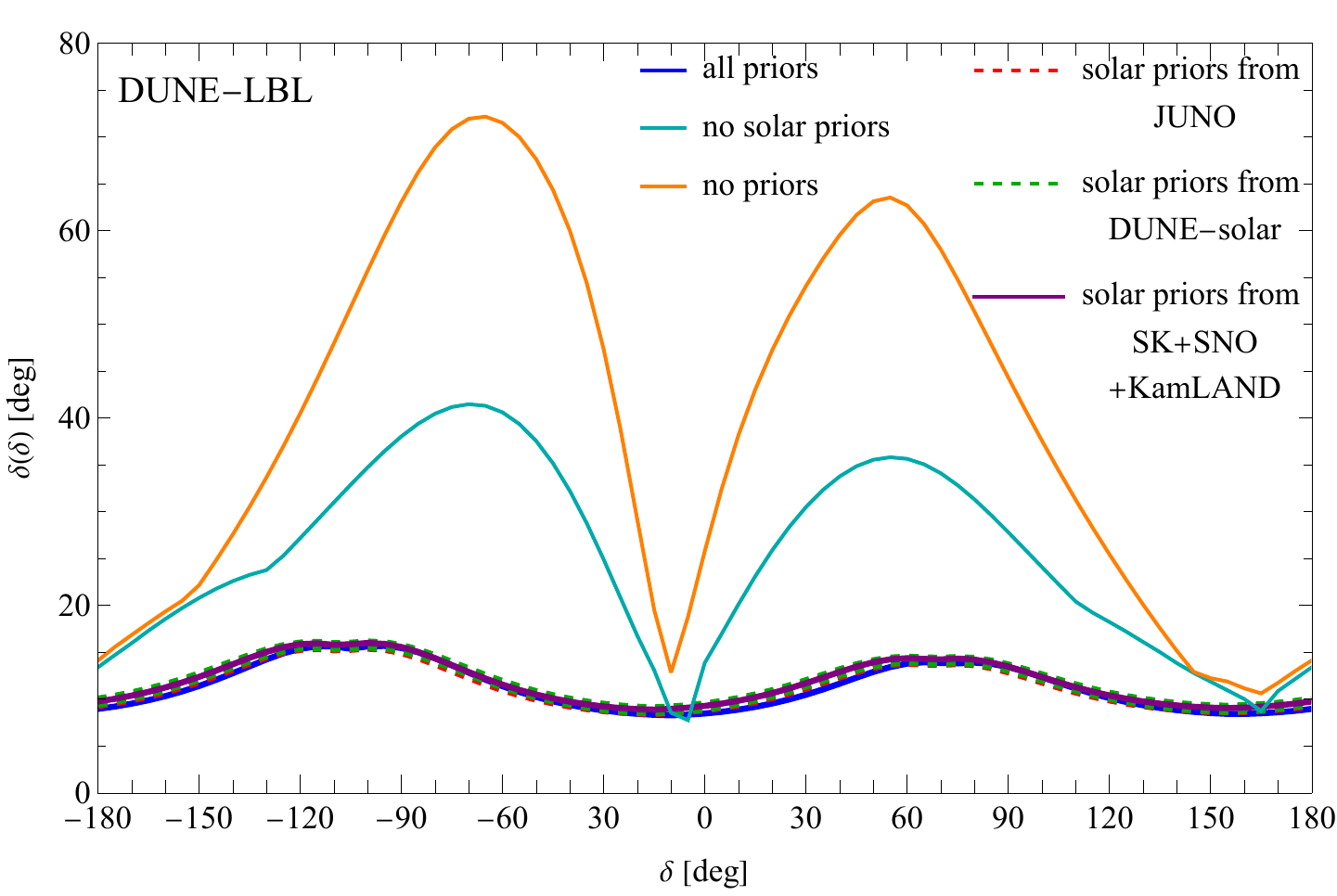}
\caption{The precision on $\delta$ at DUNE-LBL assuming the benchmark scenario defined in tab.~\ref{tab:priors} and in the text, and  different solar priors from tab.~\ref{tab:solar bf} (red, green, purple lines) and NO. 
We define the maximum of the two 1$\sigma$ uncertainties as the precision as they are asymmetric.
The blue curve shows the precision using priors on all parameters, the cyan curve shows the precision without solar priors,  and for the orange curve we do not assume any priors at all.
The red, green and purple curves  lay basically on top of each other and on top of the blue curve which assumes priors from our current knowledge of the solar parameters.}
\label{fig:precision}
\end{figure}

While determining if $\sin\delta=0$ can be disfavored or not is a crucial part of the neutrino physics program, it is also important to measure the value of $\delta$ precisely  regardless of whether it is close to CP conserving values or not to have a chance of solving the flavor puzzle \cite{Gehrlein:2022nss}.
We therefore show in fig.~\ref{fig:precision} the $1\sigma$ precision with which DUNE-LBL will be able to determine the value of $\delta$ as a function of the true value of $\delta$ with the different priors on solar experiments defined in tab.~\ref{tab:solar bf}.
For the remaining parameters we used the benchmark priors provided in the text.
We use as measure of the precision $\delta(\delta)$ defined as the  maximum of the two 1$\sigma$ uncertainties as they are often asymmetric.
This figure shows that with priors on $\Delta m^2_{21}$ and $\theta_{12}$ from current or future experiments (including those DUNE is expected to get from measuring solar neutrinos itself), DUNE-LBL can determine $\delta$ to within $\sim10^\circ-15^\circ$ precision.
Without priors on $\Delta m^2_{21}$ and $\theta_{12}$ but with priors on $\theta_{13}$ the precision is much worse, reaching  $\sim35^\circ-40^\circ$ precision around $\delta\sim\pm60^\circ$.
Finally, with no external priors at all, DUNE-LBL may only be able to determine $\delta$ to $\sim60^\circ-70^\circ$ precision.
The qualitative results for HK-LBL are similar, note however that the precision at HK-LBL worsens if the mass ordering is not fixed.

\begin{figure}
\centering
\includegraphics[width=0.65\textwidth]{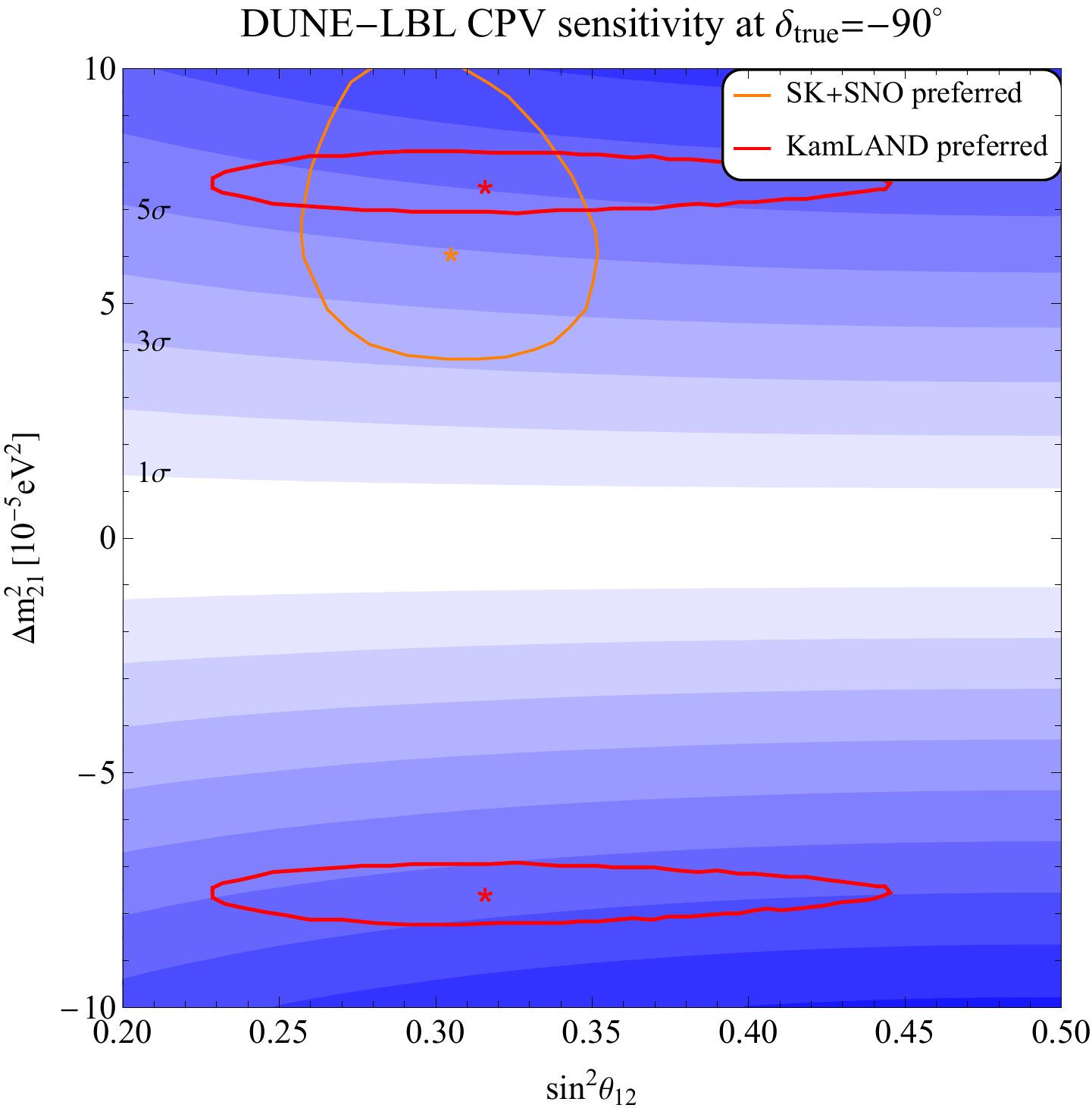}
\includegraphics[width=0.65\textwidth]{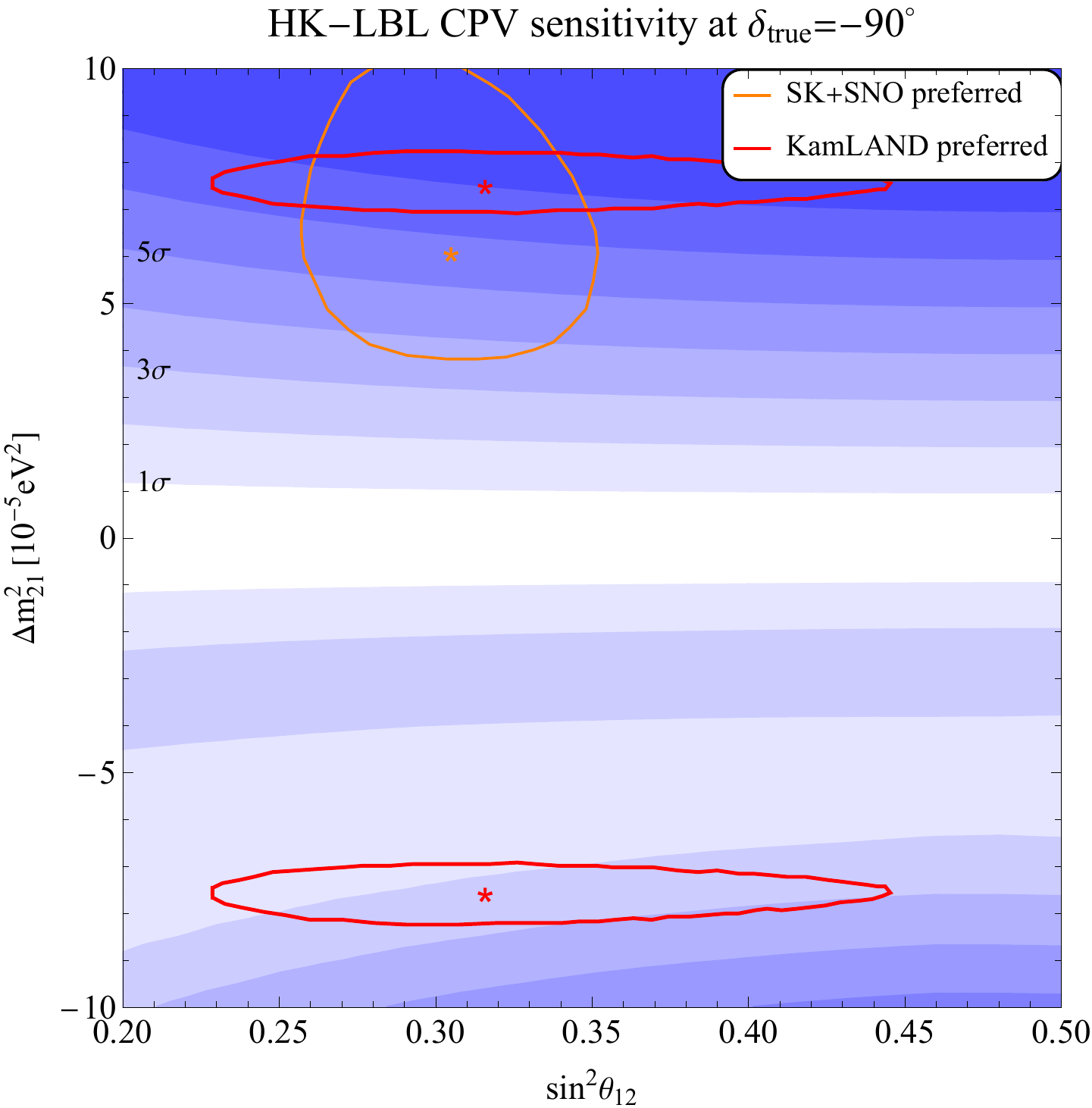}
\caption{
The sensitivity to discover CPV at $\delta=-90^\circ$ at DUNE-LBL (top) and HK-LBL (bottom) in NO while varying the true values of $\Delta m^2_{21}$ and $\theta_{12}$ but keeping their absolute uncertainty $\delta x$ fixed to the  latest combined fit of SK+SNO+KamLAND, see table \ref{tab:priors}. For the remaining parameters we use the best fit values from our benchmark case but we do not assume any priors on them.
For comparison we show the current experimental preferred regions at 3$\sigma$ for $\Delta m^2_{21}$ and $\theta_{12}$ from SK+SNO, and KamLAND using priors from reactor experiments on $\theta_{13}$.
The results assuming the true value is $\delta=90^\circ$ are shown in fig.~\ref{fig:solar2d_dune}.}
\label{fig:2d central}
\end{figure}

Now that it is clear that our knowledge of $\Delta m^2_{21}$ and $\theta_{12}$ plays a key role on our ability to measure the complex phase $\delta$, we investigate the sensitivity to $\delta$ as a function of both the precision of those priors as well as the central values.
We show in fig.~\ref{fig:2d central} how the sensitivity DUNE-LBL and HK-LBL to discover CPV at $\delta=-90^\circ$ depends on the true central values while keeping the absolute uncertainty $\delta x$ fixed to the one from SK+SNO+KamLAND. For the remaining parameters we used the best fit from our benchmark scenario but assumed no uncertainty on them.
We also show the currently preferred values for $\Delta m^2_{21}$ and $\theta_{12}$ from solar data only and from KamLAND as useful benchmarks.
We see that changing the true value of $\Delta m^2_{21}$ and $\theta_{12}$ from the best fit from KamLAND to that from solar data reduces the peak sensitivity at DUNE-LBL and HK-LBL to CP violation by $>1\sigma$.
We also see that, consistent with expectations based on the discussion in \ref{sec:solar overview}, smaller values of $\Delta m^2_{21}$ and $\theta_{12}$ lead to lower peak sensitivities to CP violation. Similar conclusions also apply when using $\delta=-90^\circ$ as true value, see fig.~\ref{fig:solar2d_dune}.
At HK-LBL negative true values of $\Delta m_{21}^2$ and $\delta=-90^\circ$ lead to a lower sensitivity to discover $\delta=-90^\circ$ compared to positive values of $\Delta m_{21}^2$ as a change of sign of $\Delta m_{21}^2$ is equivalent to a change of sign of $\Delta m_{31}^2$ where for HK degeneracies between the mass ordering and $\delta$ appear. This degeneracy is not present in DUNE as it can measure the MO at high significance.

Finally, we also test the impact of the  uncertainty of $\Delta m^2_{21}$ and $\theta_{12}$ on the CPV sensitivity. 
Unlike the impact of the central values we find that   the uncertainties only plays a minor role for the CP sensitivity. Also the impact on the precision of $\delta$ is marginal, which can also be seen from fig.~\ref{fig:precision} where the results using various priors, which differ by their uncertainty on $\Delta m^2_{21}$ and $\theta_{12}$ according to tab.~\ref{tab:priors}, are very comparable. 
We conclude that the true value of $\Delta m^2_{21}$ and $\theta_{12}$ plays a bigger role in the CPV sensitivity and precision for future experiments making a reliable knowledge of the true values highly desirable.

\subsection{Sensitivity to the solar parameters}
Next we investigate the relative role of $\Delta m^2_{21}$ and $\theta_{12}$ information in fig.~\ref{fig:mcdonalds subtract}. 
In particular, we see that with a prior on either $\Delta m^2_{21}$ or $\theta_{12}$ the sensitivity to $\delta$ is only partially degraded, but with a prior on neither solar parameter the sensitivity to $\delta$ is considerably degraded. This demonstrates that LBL experiments are sensitive to $\Delta m^2_{21}$ and $\theta_{12}$ using accelerator neutrinos, a fact previously not discussed in the literature that we are aware of.

\begin{figure}
\centering
\includegraphics[width=0.75\textwidth]{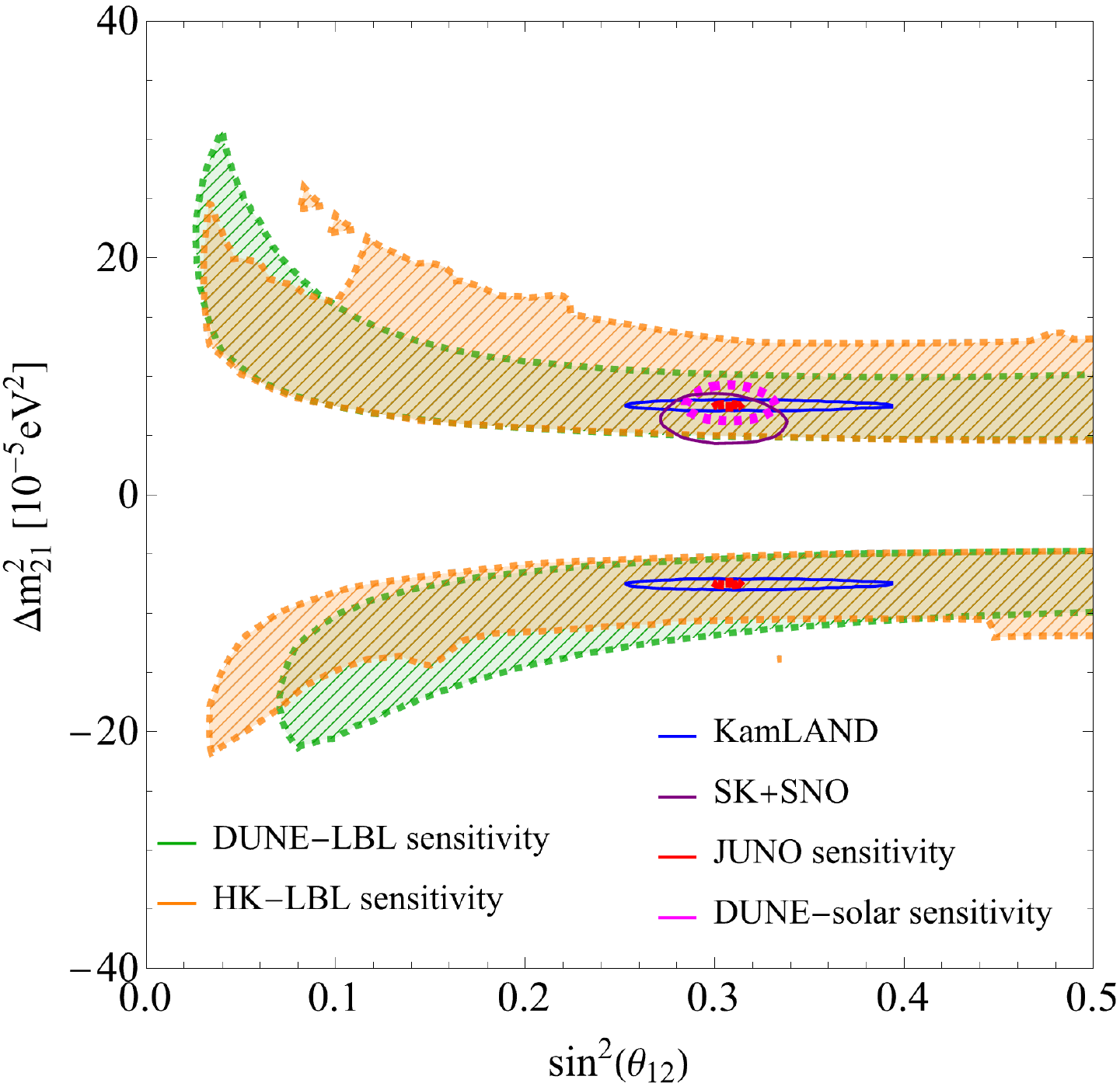}
\caption{The sensitivity to the solar oscillation parameters at upcoming long-baseline accelerator neutrino oscillation experiments  along with existing constraints from KamLAND \cite{KamLAND:2013rgu}, solar data \cite{yusuke_koshio_2022} using priors on  $\theta_{13} $ from the reactor data, and the expected future sensitivities from DUNE-solar \cite{Capozzi:2018dat,DUNE:2020ypp}, and JUNO \cite{JUNO:2022mxj}. All curves are drawn at $2\sigma$. The JUNO curve is very small due to its excellent sensitivity to $\Delta m^2_{21}$ and $\theta_{12}$.
For the other oscillation parameters we assumed priors and  the benchmark values from the text with $\delta_{true}=-90^\circ$ in NO.
The corresponding contours for T2K and NOvA extend over nearly the whole region and have been omitted for clarity.}
\label{fig:solar2d}
\end{figure}

For this reason we study the sensitivity of current and future LBL experiments using the benchmark scenario defined in sec.~\ref{sec:analysis}. In fig.~\ref{fig:solar2d} we show the sensitivity of long-baseline accelerator neutrino experiments to measure $\Delta m^2_{21}$ and $\theta_{12}$.
While current generation experiments cannot disfavor $\theta_{12}=0$ and prefer a very wide-range of $\Delta m^2_{21}$ values and are therefore omitted from this figure, DUNE-LBL and HK-LBL can measure these parameters with some precision. In fact, current LBL experiments only have very weak sensitivity to both $\Delta m^2_{21}$ and $\theta_{12}$ and allow $\Delta m_{21}^2=0$ and $\theta_{12}=0$ at the $2\sigma$ level, but future LBL experiments can exclude zero values of both $\Delta m^2_{21}$ and $\theta_{12}$ at high significance matching the ability to discover CP violation with no priors on $\Delta m^2_{21}$ and $\theta_{12}$ since if either of these parameters goes to zero, then CP is conserved.
By a similar argument, at different values of $\delta$, the precision on $\Delta m^2_{21}$ and $\theta_{12}$ will worsen.
However they cannot determine the sign of $\Delta m_{21}^2$ at high significance such that there are two disjoint preferred regions.

It may seem somewhat unexpected that DUNE-LBL does slightly better on measuring $\Delta m^2_{21}$ and $\theta_{12}$, while removing those parameters has a bigger impact on its ability to measure CPV than for HK as shown in fig.~\ref{fig:mcdonalds subtract}.
This can be understood from fig.~\ref{fig:osciprob} which shows that the variation in the probability due to $\Delta m^2_{21}$ and $\theta_{12}$ is generally as big or larger than that due to $\delta$.
Thus the remaining sensitivity to CPV or similarly to measure $\Delta m^2_{21}$ and $\theta_{12}$ must come from a combination of shape effects and neutrino/antineutrino modes.

\begin{figure}
\centering
\includegraphics[width=0.67\textwidth]{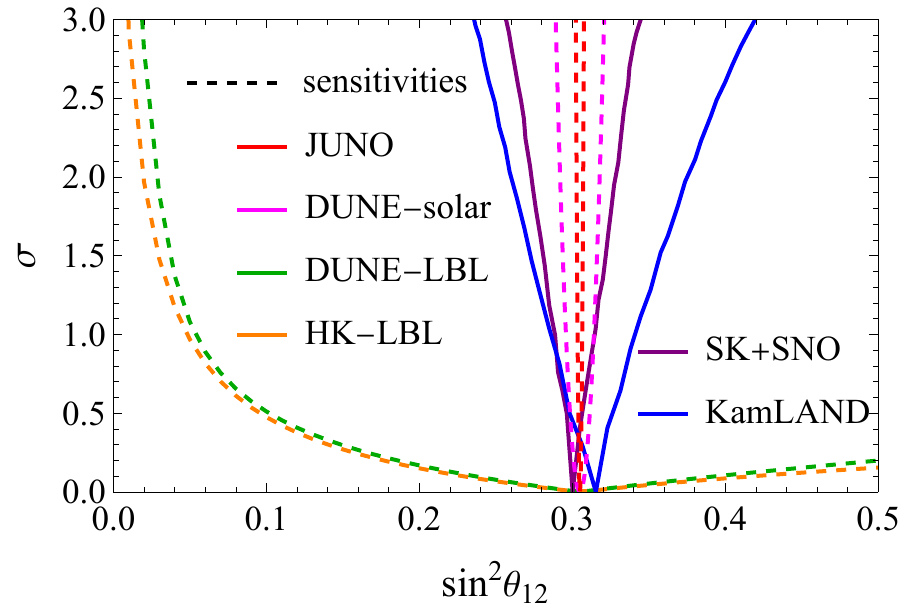}
\includegraphics[width=0.67\textwidth]{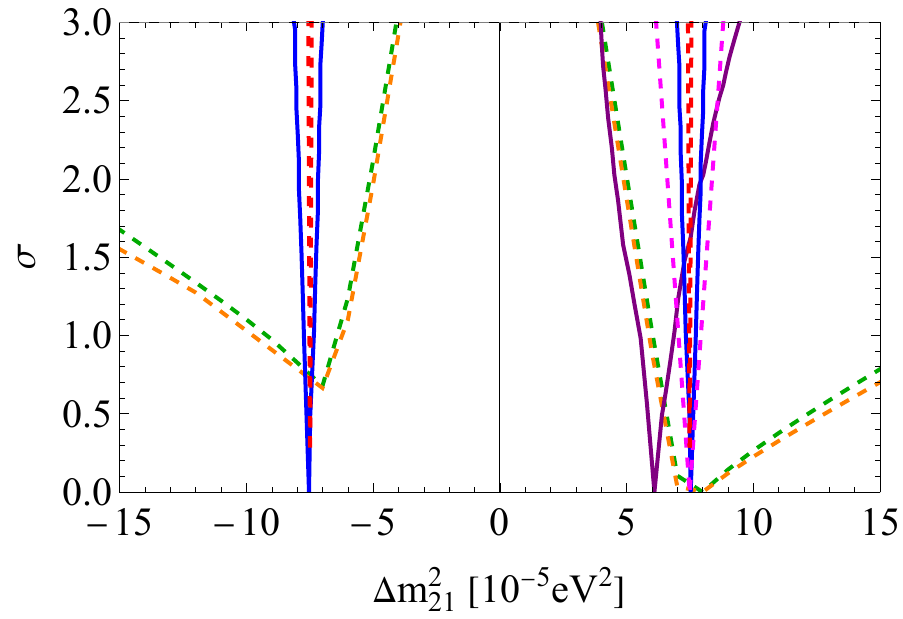}
\caption{The sensitivity of various different experiments for $\Delta m^2_{21}$ and $\theta_{12}$.
The lines use priors on $\theta_{23},~\theta_{13},~\Delta m_{31}^2 $ and the benchmark values from sec.~\ref{sec:solar overview} and $\delta_{true}=-90^\circ$ in NO. The top plot shows the sensitivities to $\theta_{12}$ and the bottom plot shows the sensitivities to the solar $\Delta m^2$. We also show the current constraints from KamLAND \cite{KamLAND:2013rgu}, solar data \cite{yusuke_koshio_2022}, and the expected future sensitivities from DUNE-solar \cite{Capozzi:2018dat,DUNE:2020ypp}, and JUNO \cite{JUNO:2022mxj}.
We do not show the sensitivities of NOvA and T2K as they only provide weak constraints on these parameters.}
\label{fig:2Dmsqs}
\end{figure}

We also point out that if we take a reasonable prior on $\theta_{12}$, we see that DUNE-LBL and, to a lesser extent, HK-LBL, can determine $\Delta m^2_{21}$ with precision comparable to within a factor of $\sim2$ of current solar measurements.
Thus, given current data and a future LBL measurement alone, the LBL measurement would provide a relevant constraint on $\Delta m^2_{21}$.
The previously mentioned caveat about the true value of $\delta$ still applies, however. We demonstrate this result in fig.~\ref{fig:2Dmsqs} 
where we show the one-dimensional constraints from long-baseline accelerator neutrino oscillation experiments on both $\Delta m^2_{21}$ and $\theta_{12}$.

We assume as true value $\delta=-90^\circ$. 
As is already discussed for  fig.~\ref{fig:solar2d} the current generation of LBL accelerator experiments is not very sensitive to $\Delta m^2_{21}$ and $\theta_{12}$.
NOvA and T2K can exclude $\sin^2\theta_{12}=0$ at 1.5-2$\sigma$ and allow  maximal $\theta_{12}$ at less than 0.5$\sigma$ and are therefore omitted from fig.~\ref{fig:2Dmsqs}.
Future LBL experiments however will provide  better constraints, DUNE-LBL and HK-LBL exclude $\sin^2\theta_{12}=0$ at 7$\sigma$ but allow maximal solar mixing at less then 0.5$\sigma$.
For $\delta=0$ the sensitivity decreases such that current LBL experiments allow all values of $\theta_{12}$, including $\theta_{12}=0$ at $\lesssim 0.5\sigma$  while future experiments can exclude $\theta_{12}=0$ at $1-2\sigma$ while they can only distinguish non-zero values with $0.5-1\sigma$ sensitivities and prefer a wide range of values for $\sin^2 \theta_{12}$.

Regarding the solar mass splitting,  current LBL experiments can provide only very mild bounds; T2K and NOvA have sensitivity to constrain $|\Delta m_{21}^2|\lesssim 45 ~\text{eV}^2$ at 3$\sigma$ and both allow $\Delta m_{21}^2=0$ at $2\sigma$. However future LBL experiments will provide slightly stronger constraints with  $ |\Delta m_{21}^2|\lesssim 35 ~\text{eV}^2$ at $3\sigma$ at HK-LBL and a slightly narrower constraint at DUNE-LBL due to its ability to measure the octant of $\theta_{23}$ with higher sensitivity.  
Both experiments present two minima at $\pm |\Delta m_{21}^2|$ where the minimum for negative $\Delta m_{21}^2$ is lifted at the $\lesssim1\sigma$ level.
This means that both future LBL experiments can also determine the sign of the solar mass splitting with some significance. 
The presence of two disparate minima also demonstrates that both experiments can exclude $\Delta m_{21}^2=0$ at a high significance ($\gtrsim 7\sigma$).

For $\delta=0$, on the other hand, the exclusion of $\Delta m_{21}^2=0$ persists due to the shape information which is stronger for DUNE than for HK but it shrinks to $4\sigma$ (DUNE-LBL) and $2\sigma$ (HK-LBL) and for current experiments to below $1\sigma$.
However the negative solution remains lifted at $1\sigma$ for the future experiments.

Furthermore, DUNE-LBL's and HK-LBL's sensitivity to $\Delta m_{21}^2$
demonstrates in particular that they can measure both dominant frequencies.
That is, they can determine the atmospheric $\Delta m^2$ whose information comes dominantly from their disappearance channel, and is thus well described by \cite{Nunokawa:2005nx}
\begin{align}
\Delta m^2_{\mu\mu}&=s_{12}^2\Delta m^2_{31}+c_{12}^2\Delta m^2_{32}+\cos\delta s_{13}\sin2\theta_{12}\tan\theta_{23}\Delta m^2_{21}\,,\\
&\approx s_{12}^2\Delta m^2_{31}+c_{12}^2\Delta m^2_{32}\,.
\end{align}
The ability to measure both dominant frequencies means that these experiments can measure all 6 oscillation parameters on their own, without any further input which provides an important test of the three-flavor oscillation picture.
There is some sensitivity to the solar mass ordering (the sign of $\Delta m^2_{21}$) due to several phenomena including the second oscillation maxima.

In comparison to current  and future constraints on $\Delta m_{21}^2$  from KamLAND and SK+SNO and JUNO  and  DUNE-solar  future LBL experiments are unlikely to improve the constraints, nevertheless they provide an important complementarity to other measurements using neutrinos from different sources and energies and therefore a crucial consistency check of the three-flavor picture.
This is very important given the modest spread in the preferred values of $\Delta m^2_{21}$ and $\theta_{12}$ from the global fits as shown in fig.~\ref{fig:gf compare}.

\section{Discussion}
\label{sec:discussion}
The quest for leptonic CP violation is one of the main targets of current and upcoming LBL experiments. However, typically all studies and analyses rely on the input of external parameters which come with their own unique systematic uncertainties which may be correlated with a LBL experiment's systematic uncertainties or may be completely unrelated.
In this context, the role of the input of external data on $\Delta m^2_{21}$ and $\theta_{12}$ from solar and long-baseline reactor experiments has not been carefully discussed in the literature yet. In this manuscript we have studied for the first time the effect of solar priors on the CPV sensitivity and precision of $\delta$ at current and upcoming LBL experiments.
We have shown that priors on $\Delta m^2_{21}$ and $\theta_{12}$ are quite relevant for the sensitivity to CP violation but, quite interestingly, LBL accelerator experiments still have some sensitivity to CP violation even with no information from the solar parameter experiments.
This then implies that LBL accelerator experiments have sensitivity to $\Delta m^2_{21}$ and $\theta_{12}$ at a certain level.

To better understand the impact of these priors on the measurement of $\delta$ we note that in order to reduce the sensitivity to $\delta$ for large $|\sin\delta|$ one needs solar parameters highly inconsistent with existing data.
In particular, we find that the data would be most comparable to a CP conserving $\sin\delta=0$ scenario for $\Delta m^2_{21}\simeq60\times10^{-5}$ eV$^2$ instead of the usual $7.5\times10^{-5}$ eV$^2$.
This alone is not enough, one also needs $\theta_{12}\sim10^\circ$ for $\delta$ not too near $0,\pi$.
It turns out that the significant change to either solar parameter alone does not do a good job; only both of them together are able to approximately mimic a CP conserving scenario.
This is why relaxing the solar input to only one of the parameters does not significantly reduce the sensitivity to CP violation.

We also notice that the matter effect and the ability to measure the atmospheric mass ordering plays a key role.
For example, DUNE-LBL will have excellent sensitivity to the atmospheric mass ordering, even without priors on $\Delta m^2_{21}$ and $\theta_{12}$, because DUNE-LBL is the only experiment that will measure the mass ordering at high significance via the matter effect in $\Delta m^2_{31}$ oscillations (all measurements of the atmospheric mass ordering require a measurement of the matter effect somewhere), DUNE-LBL is also able to determine the solar mass ordering  (that is, that $\Delta m^2_{21}>0$) at $\sim1\sigma$.
HK-LBL has comparable sensitivity to the solar mass ordering since it will determine the atmospheric mass ordering at $>1\sigma$ and knowledge of the atmospheric mass ordering is a prerequisite to determining the solar mass ordering at LBL experiments.

Finally, for completeness, we briefly comment on the role priors on $\Delta m^2_{21}$ and $\theta_{12}$ as well as $\theta_{13}$ will have on the determination of the other oscillation parameters, $\Delta m^2_{31}$, $\theta_{23}$, and $\theta_{13}$.
We show several plots in appendix \ref{sec:other_para} and discuss the results here.
Some versions of these questions have been asked before, see e.g.~\cite{Fogli:1996pv,Barger:2001yr,Chatterjee:2013qus}, but not in terms of the role of $\Delta m^2_{21}$ and $\theta_{12}$.
In the appendix we show the sensitivities of current and future LBL experiments to $\theta_{23}$, $\Delta m_{31}^2$, $\theta_{13}$ with all priors and without priors on key combinations of the $\Delta m^2_{21}$, $\theta_{12}$, and $\theta_{13}$.
We find the impact of the solar priors on the sensitivity to $\theta_{23}$ is minor and most pronounced for T2K.
The lack of a prior on $\Delta m_{21}^2$ is responsible for a significant portion of this.
Even more important than solar priors is the prior on $\theta_{13}$ which we show affects the octant sensitivity more that the priors on $\Delta m^2_{21}$ and $\theta_{12}$.

Upcoming experiments will not improve Daya Bay's measurement of $\theta_{13}$ but without solar priors, the sensitivity is further reduced such that considerably smaller values of $\theta_{13}$ are allowed. We show that without a prior on $\Delta m_{21}^2$, the sensitivity declines making this the most important prior for the determination of $\theta_{13}$ at LBL experiments. Without priors on both $\Delta m^2_{21}$ and $\theta_{12}$, the allowed ranges extend to smaller values of $\theta_{13}$, in the case of T2K $\theta_{13}=0$ is allowed at $3\sigma$.

LBL experiments also have good sensitivity to $\Delta m_{32}^2$ in the disappearance channel.
In order to derive the constraint on $\Delta m_{31}^2$, however knowledge on $\Delta m_{21}^2$ is required since the measurement depends on a weighted combination of $\Delta m^2_{31}$ and $\Delta m^2_{32}$, therefore the sensitivity to $\Delta m_{31}^2$ is substantially reduced without solar priors.
In particular, without a prior on $\Delta m_{21}^2$ the effect is considerable for current LBL experiments but the effect is also significant for future experiments. In fact, a second nearly degenerate minimum appears around $\Delta m_{31}^2\approx2.42\times 10^{-3}~\text{eV}^2$ which corresponds to the resulting value of $\Delta m_{31}^2$ from the measurement of $\Delta m_{32}^2$ but with $-\Delta m_{21}^2$.
Finally without priors on solar data HK-LBL will not be able to measure the atmospheric mass ordering at more than $\sim 1\sigma$ and HK-LBL's precision on $\Delta m^2_{31}$ gets considerably worse; the same effect is present, and even more dramatic, for the current LBL experiments.

\section{Conclusion}
\label{sec:conclusion}
We have demonstrated that solar oscillation parameters, $\theta_{12}$ and $\Delta m^2_{21}$, play an important and largely unrecognized role in long-baseline (LBL) neutrino oscillations.
In particular, without external knowledge on $\Delta m^2_{21}$ and $\theta_{12}$, LBL experiments have significantly limited sensitivity to discover CP violation.
Moreover, $\Delta m^2_{21}$ and $\theta_{12}$ can actually be determined at LBL experiments providing a valuable cross check, especially given that there a spread of preferred values from the experiments and the global fits.
In addition, the true values for the solar parameters, in particular $\Delta m^2_{21}$, plays an important role in the ability to measure CPV with LBL neutrinos affecting the peak sensitivity to CPV by $>1\sigma$.
Thus having precise measurements of $\Delta m^2_{21}$ and $\theta_{12}$ from e.g.~JUNO is important for determining CPV.

We also found that $\Delta m^2_{21}$ and $\theta_{12}$ have an impact on the determination of other oscillation parameters like $\theta_{13}$ and $\Delta m_{31}^2$ and  the octant of $\theta_{23}$, making precise knowledge of $\Delta m^2_{21}$ and $\theta_{12}$ fundamental for the success of the next generation of LBL experiments. In turn the sensitivity of LBL accelerator experiments to $\Delta m^2_{21}$ and $\theta_{12}$ allows us to probe the three-flavor paradigm in one experiment only and provides a consistency check of our understanding of neutrino oscillations in a different energy and baseline regime than usually used to determine $\Delta m^2_{21}$ and $\theta_{12}$.

\begin{acknowledgments}
We thank the anonymous referee for helpful comments.
We thank Francesco Capozzi for sending us the data release from \cite{Capozzi:2021fjo}.
The authors acknowledge support by the United States Department of Energy under Grant Contract No.~DE-SC0012704.
\end{acknowledgments}

\appendix

\section{Additional figures}
\label{sec:add_fig}
To complement the figures in the main text, we performed various additional parameter scans to further elucidate the interplay of $\Delta m^2_{21}$ and $\theta_{12}$ in current and next-generation long-baseline neutrino oscillation experiments. 

\begin{figure}
\centering
\includegraphics[width=0.65\textwidth]{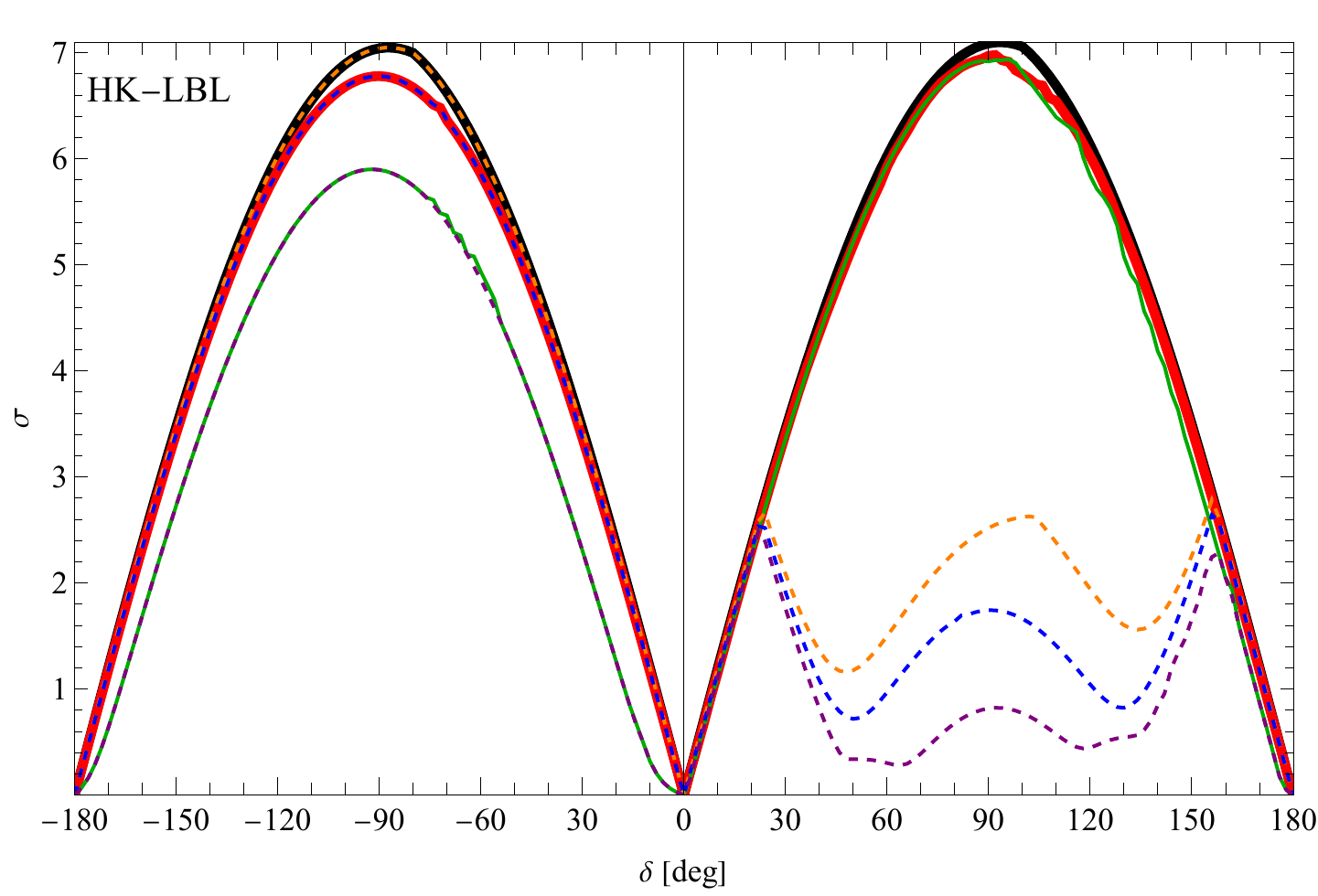}
\includegraphics[width=0.65\textwidth]{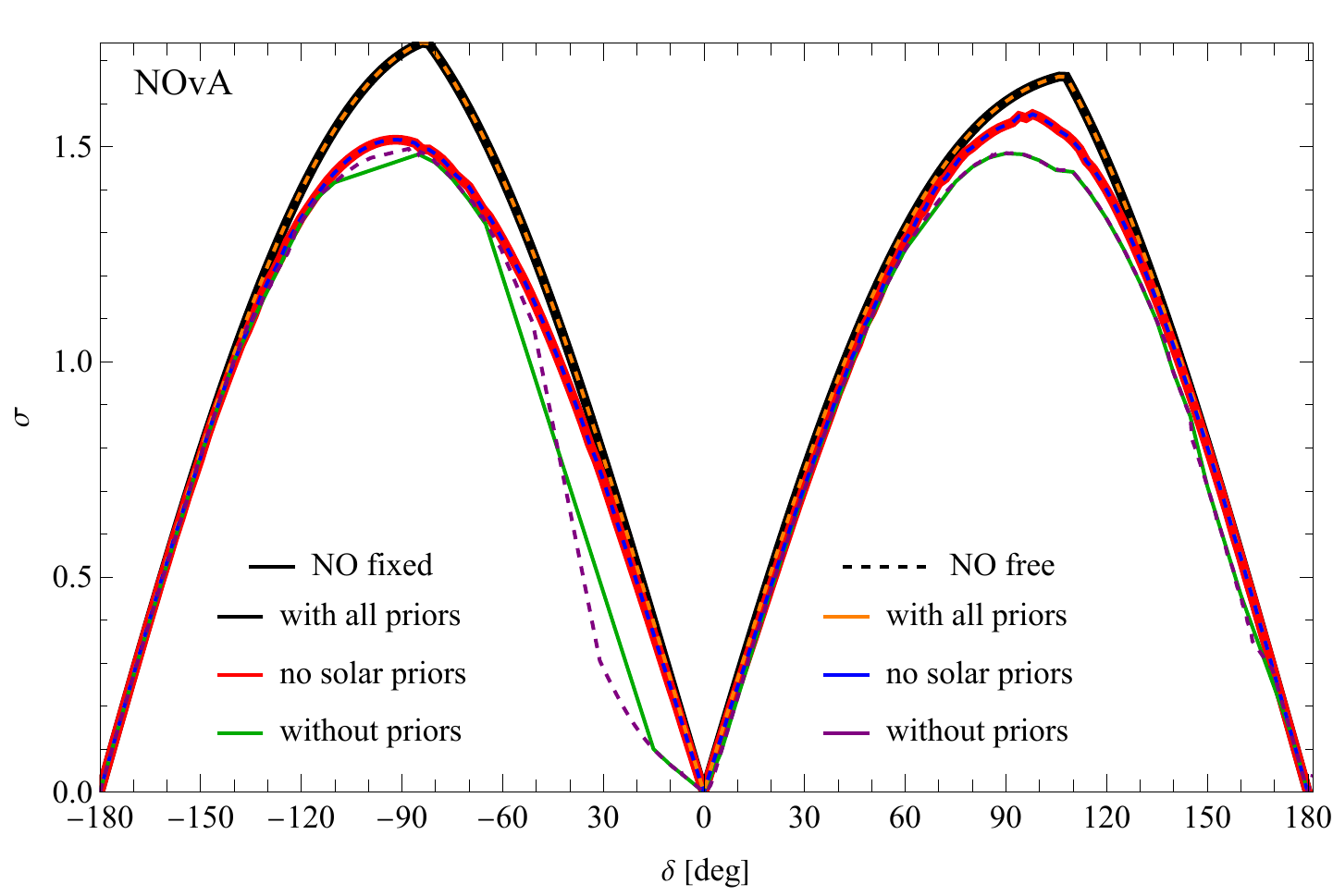}
\includegraphics[width=0.65\textwidth]{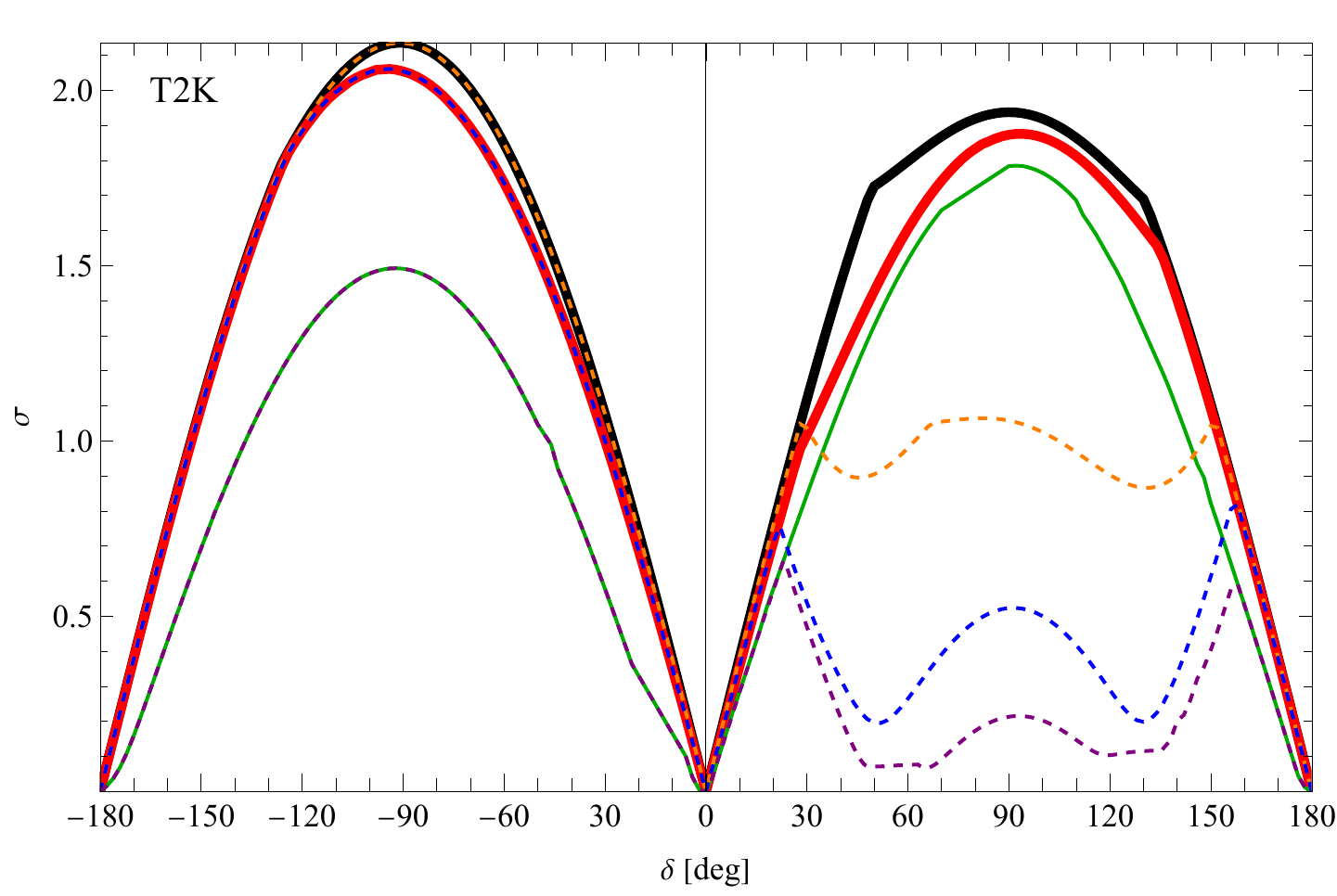}
\caption{The sensitivities to disfavor $\sin\delta=0$ of HK-LBL, NOvA, and T2K corresponding to the exposures from tab.~\ref{tab:exps} with fixed normal ordering or free  mass ordering. The black and orange lines shows the results using all priors, red and blue lines are the results without solar priors while the green and purple lines show the results without any priors.
DUNE is not shown here due to its excellent sensitivity to the MO, thus fixing the MO does not have an impact on the sensitivity to disfavor $\sin\delta=0$.}
\label{fig:hk_seni}
\end{figure}

Fig.~\ref{fig:hk_seni} shows the CPV sensitivities of HK-LBL, NOvA, and T2K under the assumption of different priors and fixed or free mass ordering. 
Note that the for HK-LBL and T2K without external information on the mass ordering, in the true NO it is very hard to discover CP violation for $\delta\in[0^\circ,180^\circ]$ as $\Delta m^2_{21}$ and $\theta_{12}$ can radically change. Note that we do not show the corresponding plot for DUNE due to its excellent sensitivity to the MO, thus fixing the MO does not have an impact on the sensitivity to
disfavor $\sin \delta = 0$.

\begin{figure}
\centering
\includegraphics[width=0.67\textwidth]{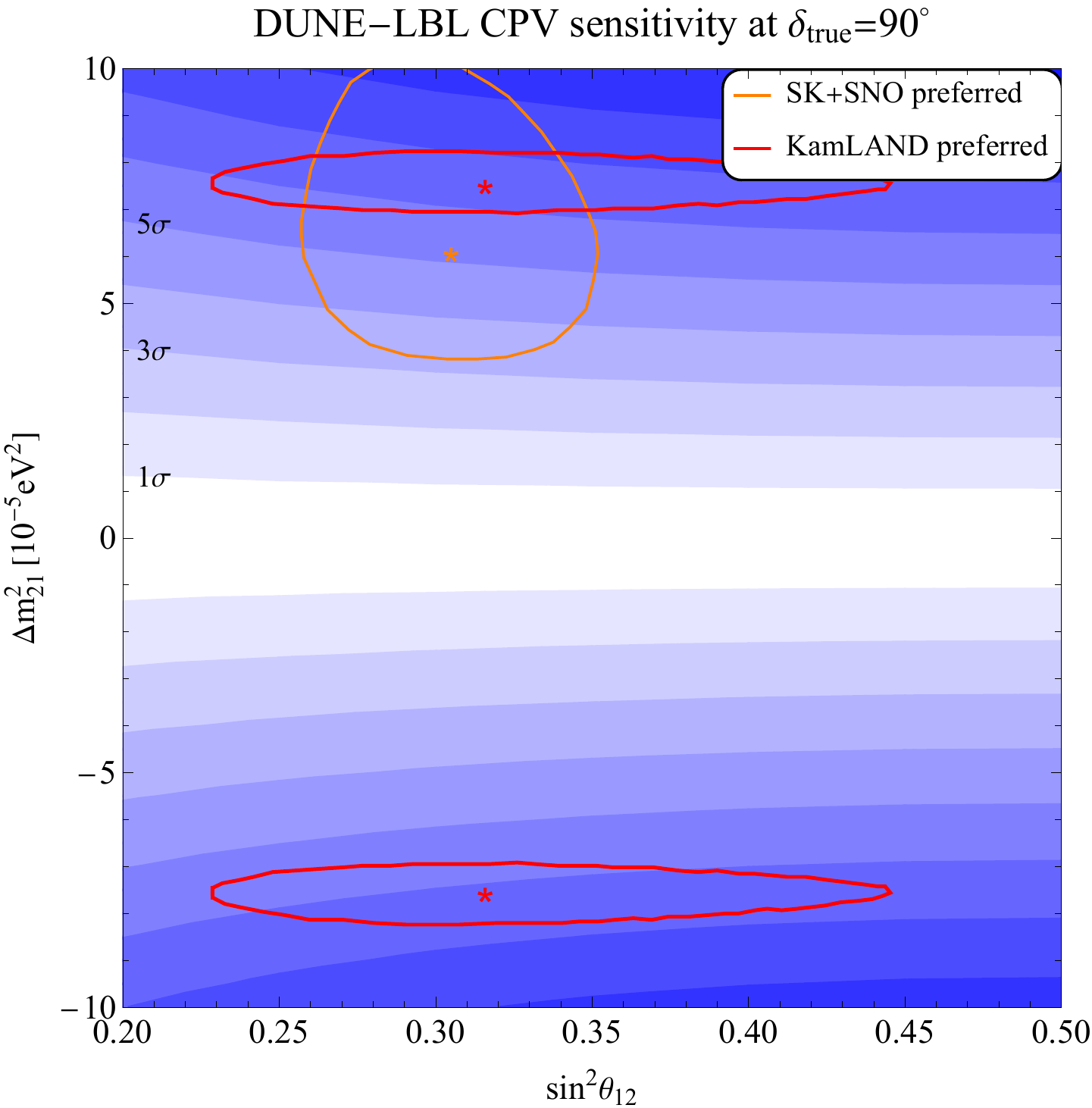}
\includegraphics[width=0.67\textwidth]{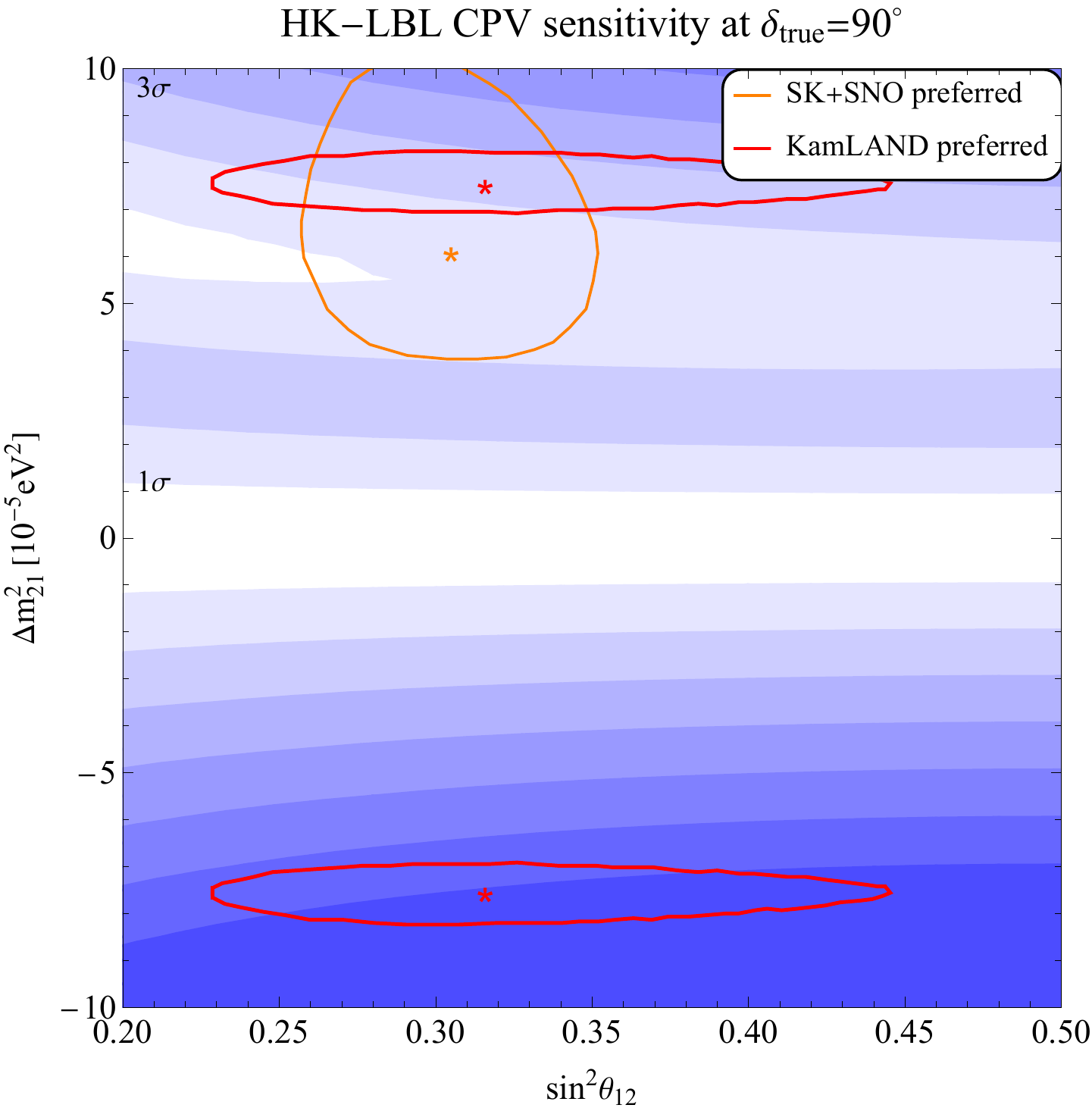}
\caption{The sensitivity to discover CPV at $\delta=90^\circ$ at DUNE-LBL (top) and HK-LBL (bottom) while varying the true values of $\Delta m^2_{21}$ and $\theta_{12}$ but keeping their absolute uncertainty $\delta x$ fixed to the  latest combined fit of SK+SNO+KamLAND, see table \ref{tab:priors}. For the remaining parameters we use the best fit values from our benchmark case but we do not assume any priors on them, we assume NO.
For comparison we show the current experimental preferred regions at 3$\sigma$ for $\Delta m^2_{21}$ and $\theta_{12}$ from SK+SNO, and KamLAND.}
\label{fig:solar2d_dune}
\end{figure}

Finally, in fig.~\ref{fig:solar2d_dune} we show the DUNE-LBL and HK-LBL sensitivities at $\delta=90^\circ$ for different central values of $\Delta m^2_{21}$ and $\theta_{12}$ keeping their absolute uncertainty $\delta x$ fixed to the one from the latest combined fit of SK+SNO+KamLAND from table \ref{tab:priors}.

\section{The impact of solar parameters on the determination of the atmospheric parameters}
\label{sec:other_para}
\begin{figure}
\centering
\includegraphics[width=0.49\textwidth]{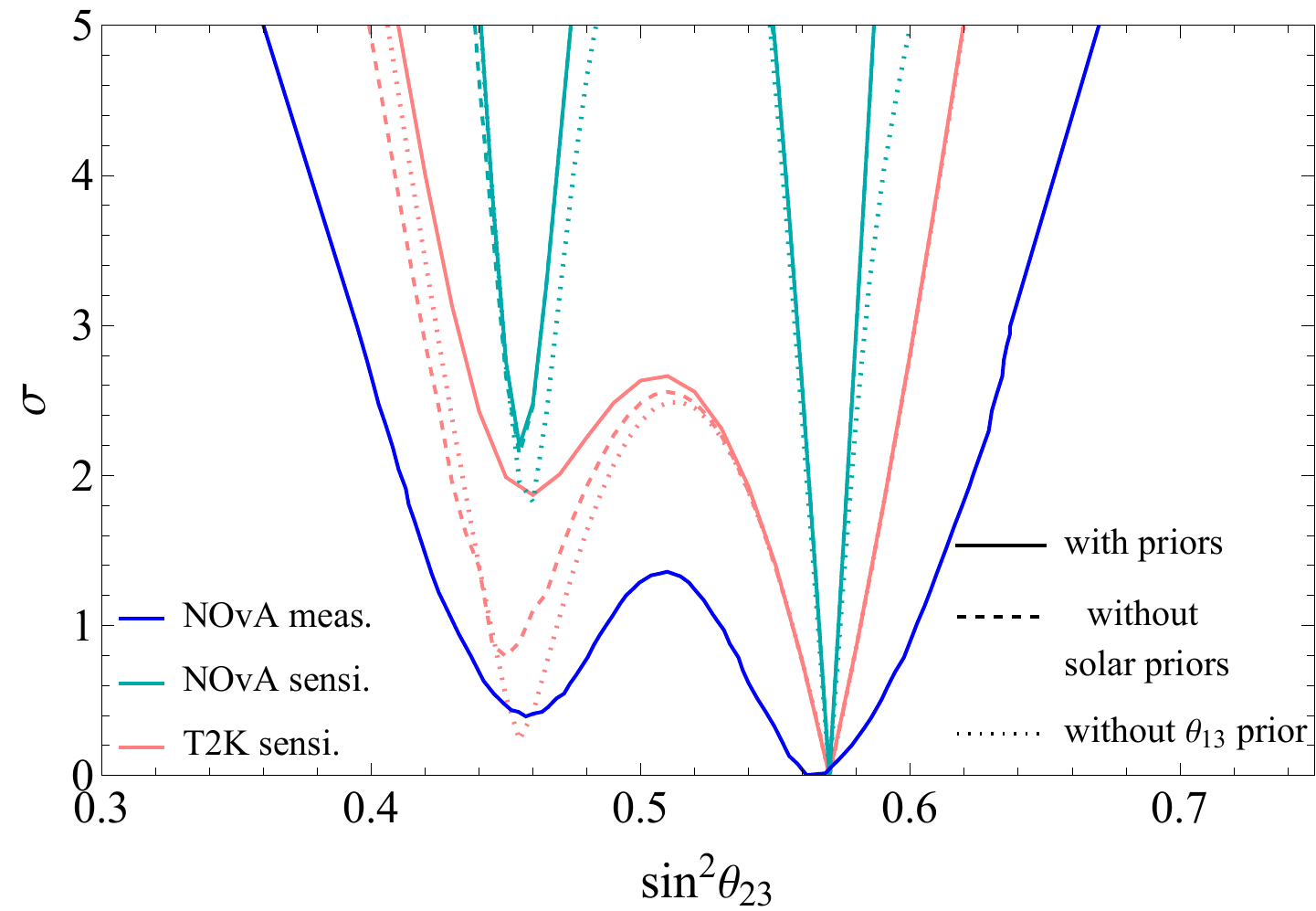}
\includegraphics[width=0.49\textwidth]{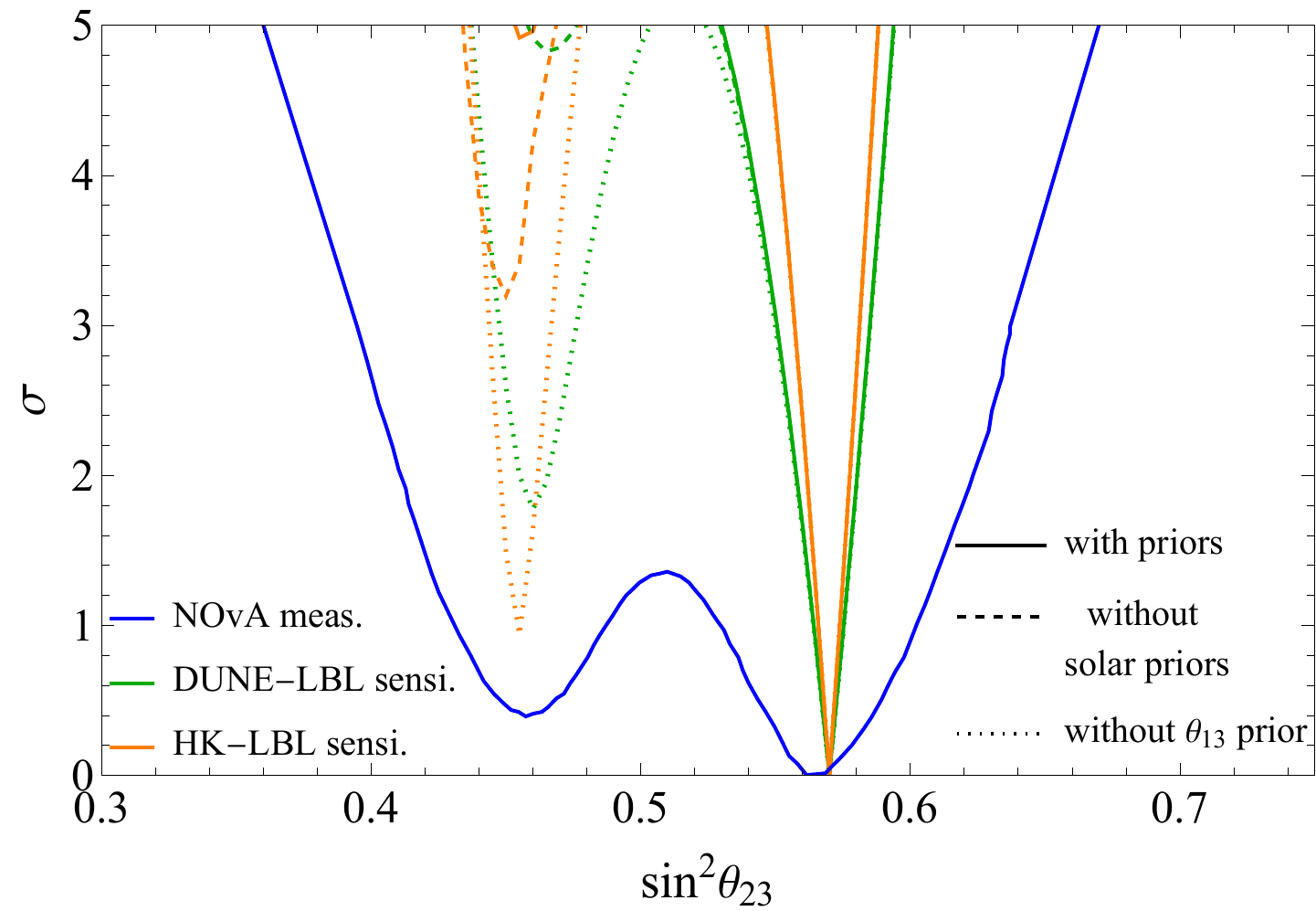}
\caption{Sensitivity to $\theta_{23}$ of current (left) and future LBL experiments (right) compared to the current best measurement from NOvA \cite{NOvA:2021nfi} in NO. The solid lines show the sensitivity using priors on all parameters but $\delta$, the dashed lines do not use solar priors. The dotted curve shows the results without a prior on $\theta_{13}$.
The solar parameters have an impact on the sensitivity to the sensitivity to the octant.  For the other parameters we assumed the benchmark values from above and $\delta_{true}=-90^\circ$.}
\label{fig:th23_sensi}
\end{figure}

\begin{figure}
\centering
\includegraphics[width=0.49\textwidth]{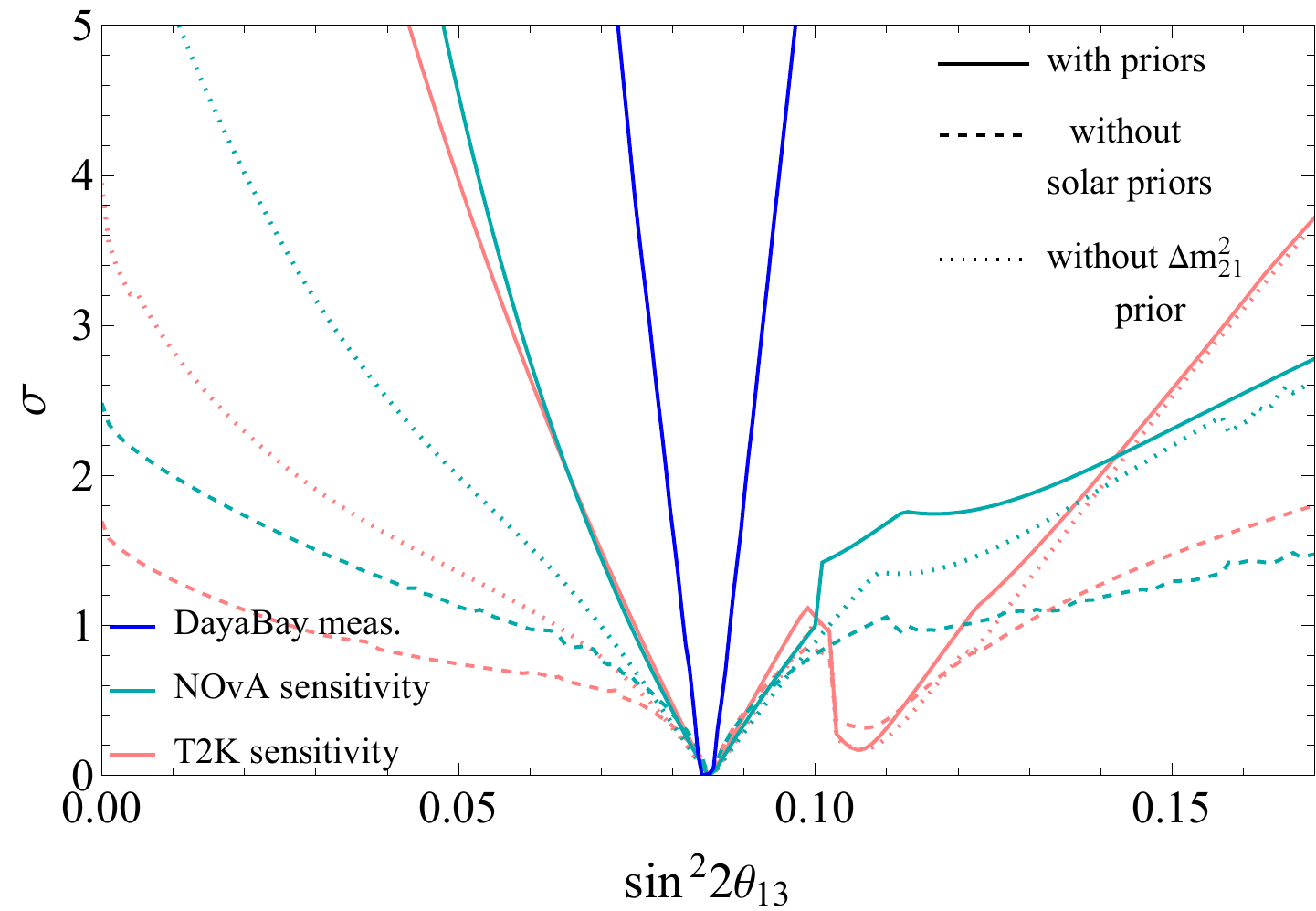}
\includegraphics[width=0.49\textwidth]{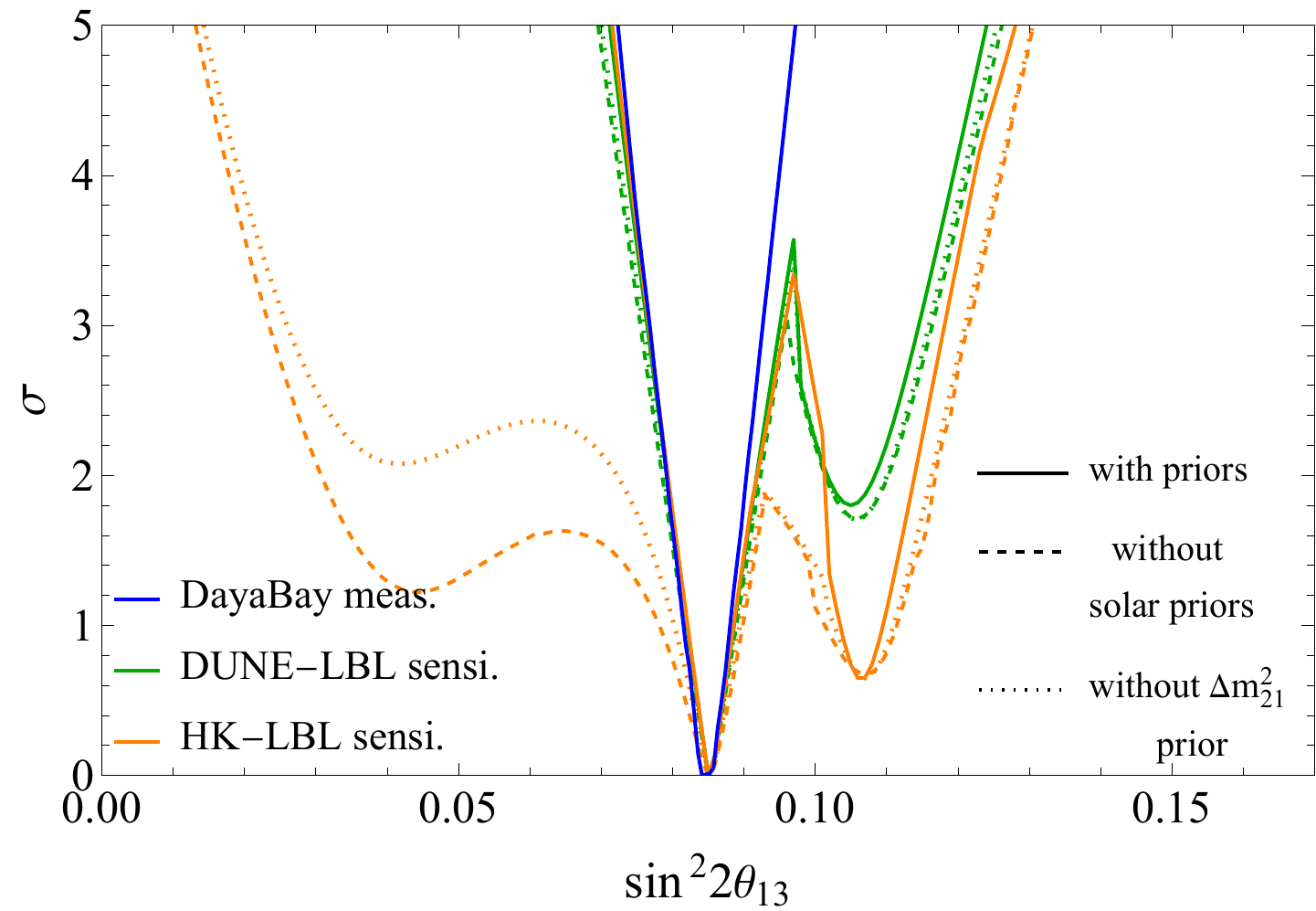}
\caption{Sensitivity to $\theta_{13}$ of current (left) and future LBL experiments (right) compared to the current best measurement from Daya Bay \cite{kam_biu_luk_2022_6683712} in NO. 
The solid lines show the sensitivity using priors on all parameters but $\delta$, the dashed lines do not use solar priors. The dotted curve shows the results without a prior on $\Delta m_{21}^2$.
The second higher local minimum comes from the other octant of $\theta_{23}$. For the other parameters we assumed the benchmark values from above and $\delta_{true}=-90^\circ$.}
\label{fig:th13_sensi}
\end{figure}

\begin{figure}
\centering
\includegraphics[width=0.49\textwidth]{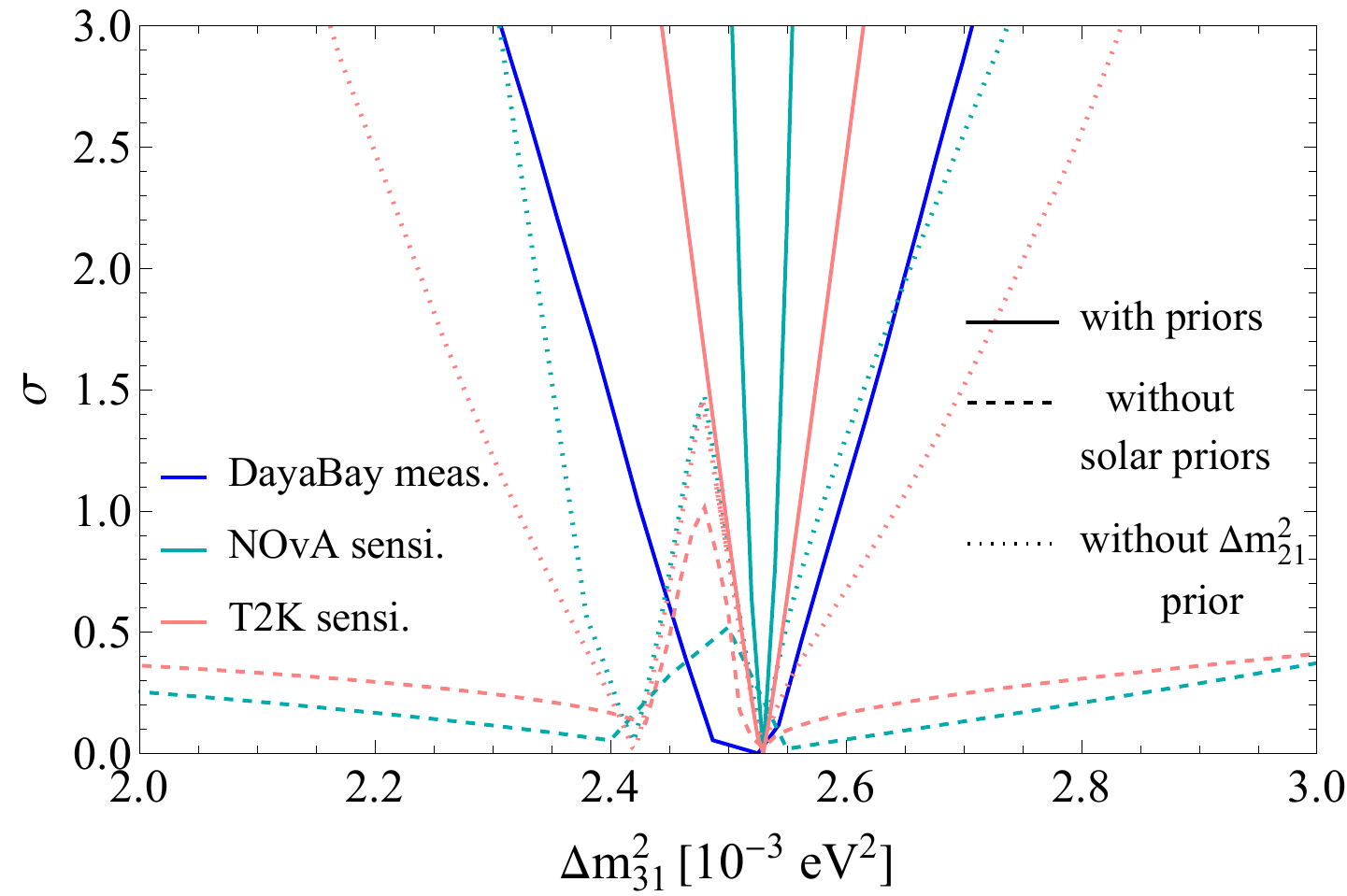}
\includegraphics[width=0.49\textwidth]{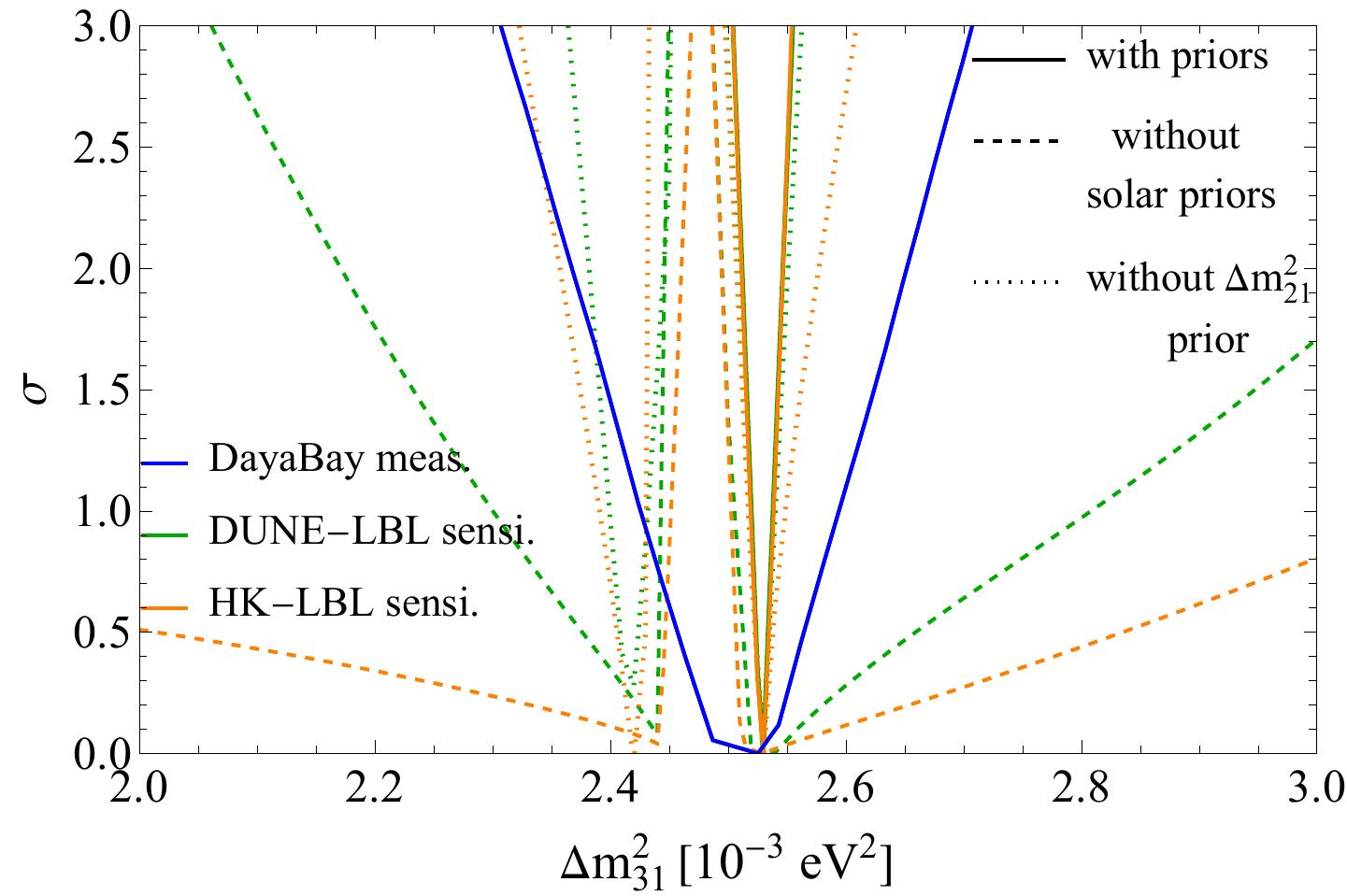}
\caption{Sensitivity to $\Delta m_{31}^2$ of current (left) and future LBL experiments (right) compared to the current best measurement from Daya Bay \cite{kam_biu_luk_2022_6683712} in NO. We show the sensitivity using priors on all parameters but $\delta$, no solar priors, and without a prior on $\Delta m_{21}^2$. As true values for the parameters we assumed the benchmark values from above and $\delta_{true}=-90^\circ$.
LBL experiments are chiefly sensitive to $\Delta m_{32}^2$ in muon disappearance and in a three-flavor scenario $\Delta m_{31}^2$ is then derived  from $\Delta m_{31}^2=\Delta m_{32}^2+\Delta m_{21}^2$. The two minima in the case without prior on $\Delta m_{21}^2$ correspond to
the different signs of $\Delta m_{21}$ while
the value $\Delta m_{31}^2\approx 2.45~\text{eV}^2$ is excluded at 4-5$\sigma$ at HK-LBL and DUNE-LBL as in this case $\Delta m_{31}^2=\Delta m_{32}^2$ and $\Delta m_{21}^2$ is required to be zero which is disfavored at HK-LBL and DUNE-LBL, see fig.~\ref{fig:2Dmsqs}. 
}
\label{fig:dm31_sensi}
\end{figure}

The solar parameters also affect the measurement of the other parameters, apart from $\delta$, which we demonstrate in figs.~\ref{fig:th23_sensi}, \ref{fig:th13_sensi}, \ref{fig:dm31_sensi}.

As DUNE-LBL and NOvA have good sensitivity to $\theta_{23}$ on their own, the impact of the absence of solar priors is very small. This is different for T2K and HK-LBL where the absence of solar priors affects their sensitivity to resolve the octant. The dominant source for the reduction of the sensitivity is the absence of the prior in $\Delta m_{21}^2$. Furthermore, we show that the prior on $\theta_{13}$ is even more important to resolve the octant.
This is because the octant information comes from  presence of the $s_{23}^2$ term in the $\nu_\mu\to\nu_e$ appearance probability which is paired up with $s_{13}^2$, see e.g.~\cite{Chatterjee:2013qus,Cervera:2000kp,Denton:2016wmg,Barenboim:2019pfp}.

Also for the $\theta_{13}$ sensitivity the absence of solar priors affects DUNE-LBL's sensitivity only marginally whereas T2K's, NOvA's and HK-LBL's sensitivities worsen and the allowed ranges extend to smaller values of $\sin^2 2\theta_{13}$. Without solar priors T2K cannot exclude $\theta_{13}=0$ at more than $3\sigma$. Also in this case the prior on $\Delta m_{21}^2$ is important.
Finally, the sensitivity to $\Delta m_{31}^2$ gets severely affected without solar priors, in particular $\Delta m_{21}^2$, at T2K, NOvA and HK-LBL, while the effect at DUNE-LBL is smaller. The reduction of sensitivity can be understood from the fact that LBL experiments are somewhat more sensitive to $\Delta m_{32}^2$ in $\nu_\mu$ disappearance and $\Delta m_{31}^2$ is then derived from the sum rule $\Delta m_{31}^2=\Delta m_{32}^2+\Delta m_{21}^2$.
As the LBL experiments' sensitivity to $\Delta m_{21}^2$ leads to less severe constraints than our current knowledge of this parameter the derived  sensitivities of $\Delta m_{31}^2$ worsen as well. In fact, two nearly degenerate minima appear which correspond to the different signs of $\Delta m_{21}^2$ to which LBL experiments are not very sensitive.
However as we have shown in fig.~\ref{fig:2Dmsqs} the LBL accelerator experiments have some sensitivity to $\Delta m^2_{21}$ and $\theta_{12}$, in particular future experiments disfavor $\Delta m_{21}^2=0$ at a high significance. This leads to  $\Delta m_{31}^2\approx2.45\times 10^{-3}~\text{eV}^2$ to be excluded  as in this case $\Delta m_{31}^2=\Delta m_{32}^2$ and $\Delta m_{21}^2$ is required to be zero.

\bibliographystyle{JHEP}
\bibliography{main}

\end{document}